\documentclass[aps,twocolumn,showpacs,amsmath,amssymb,floatfix]{revtex4}
\usepackage{young}
\usepackage{float}
\usepackage{graphicx}
\usepackage{color}
\usepackage[latin1]{inputenc} 
\usepackage{lmodern} 
\usepackage[T1]{fontenc}

\def\nn{\nonumber}

\def \bc {\begin{center}}
\def \ec {\end{center}}
\def \bi {\begin{itemize}}
\def \ei {\end{itemize}}

\def \ba {\begin{array}}
\def \ea {\end{array}}

\def \be {\begin{equation}}
\def \ee {\end{equation}}
\def \bea {\begin{eqnarray}}
\def \eea {\end{eqnarray}}

\def \um {\frac{1}{2}}
\def\tr {\mathrm{tr}}
\def\cD {{\cal D}}

\def\DZ {\Delta_\mathrm{Z}}
\def\DS {\Delta_\mathrm{t}}
\def\ic {\mathrm{i}}

\newcommand{\la}{\langle}
\newcommand{\ra}{\rangle}

\begin{document}

\title{Husimi function and phase-space analysis of bilayer quantum Hall systems at $\nu=2/\lambda$}

\author{M. Calixto and C. Pe\'on-Nieto}

\affiliation{Departamento de Matem\'atica Aplicada and Instituto ``Carlos I'' de F\'isica Te\'orica y Computacional,  Universidad de Granada,
Fuentenueva s/n, 18071 Granada, Spain}

\date{\today}

\begin{abstract}
We propose localization measures in phase space of the ground state of bilayer quantum Hall (BLQH) systems at fractional filling factors $\nu=2/\lambda$, to characterize the three quantum phases 
(shortly denoted by spin, canted and ppin) for arbitrary $U(4)$-isospin $\lambda$. We use a coherent state (Bargmann) representation of quantum states, as holomorphic functions in the 8-dimensional 
Grassmannian phase-space  $\mathbb{G}^4_{2}=U(4)/[U(2)\times U(2)]$ (a higher-dimensional generalization of the Haldane's 2-dimensional sphere $\mathbb{S}^2=U(2)/[U(1)\times U(1)]$). 
We quantify the localization (inverse volume) of the ground state wave function in phase-space throughout the phase diagram 
(i.e., as a function of Zeeman, tunneling, layer distance, etc, control parameters) with the Husimi function second moment, a kind of inverse participation ratio that behaves as an order parameter. 
Then we visualize the different ground state structure in phase space of the three quantum phases, the canted phase displaying a much higher delocalization (a Schr\"odinger cat structure) than the spin and ppin phases, where the ground 
state is highly coherent. We find a good agreement between analytic (variational) and numeric diagonalization results.

\end{abstract}


\maketitle

\section{Introduction}

Information theoretic and statistical measures have proved to be useful in the description and characterization of quantum phase transitions (QPTs). For example, 
in the traditional Anderson metal-insulator transition \cite{Anderson,Wegner,Brandes,Aulbach,Evers}, Hamiltonian eigenfunctions underlie strong fluctuations. Also, 
the localization of the electronic wave function can be regarded as the key manifestation of quantum coherence at a macroscopic scale in a 
condensed matter system. In this article we want to analyze QPTs in BLQH systems at fractional filling factors $\nu=2/\lambda$ from an information theoretic perspective. 
The integer case $\nu=2$ has been extensively studied in the literature (see e.g. \cite{HamEzawa,EzawaBook,PRB60,PRLBrey,Schliemann,Ezawabisky,fukudamagnetotransport}), 
where the analysis of the ground state structure reveals the existence of (in general) three quantum phases, shortly denoted by: spin, canted and ppin \cite{HamEzawa,EzawaBook}, 
depending on which order parameter (spin or pseudospin/layer) dominates across the control parameter space: tunneling, Zeeman, bias voltage, etc, couplings. Here we 
shall study localization properties in phase-space of the ground state in each of the three quantum phases for arbitrary $\lambda$ (number of magnetic flux quanta per electron). 
For it, we shall use a phase-space representation $\la Z|\psi\ra=\psi(Z)$ of quantum states $|\psi\ra$, where $|Z\ra$ denotes and arbitrary coherent (minimal uncertainty) state labeled by points 
$Z\in \mathbb{G}^4_{2}=U(4)/[U(2)\times U(2)]\simeq \mathrm{Mat}_{2\times 2}(\mathbb{C})$ ($2\times 2$ complex matrices), the 
complex Grassmannian with complex dimension 4. The Grassmannian $\mathbb{G}^4_{2}$ can be seen as a higher-dimensional generalization of the Haldane's sphere $\mathbb{S}^2=U(2)/[U(1)\times U(1)]$ \cite{Haldane} 
for monolayer fractional QH systems, with $Z$ a $2\times 2$ matrix generalization of the stereographic projection $z=\tan(\theta/2)e^{\ic\phi}$ of a
point $(\theta,\phi)$ (polar and azimuthal angles) of the Riemann sphere $\mathbb S^2$ onto the complex plane. Standard spin-$s$  coherent states (CS) on the sphere  $\mathbb{S}^2=\mathbb{C}P^1$ (isomorphic to the complex projective space) 
are very well known (see traditional references \cite{Klauder,Perelomovbook,Gazeaubook,Antoine} on CS and later on section \ref{secHaldane}). 
Its extension to the complex projective space $\mathbb{C}P^{N-1}=U(N)/[U(N-1)\times U(1)]$ (the symmetric case) is quite straightforward 
and $\mathbb{C}P^{N-1}$ is related to the phase space of $N$-component QH systems at fractional values of $\nu=1$. The case $1<\nu<N/2+1$ is much more involved and the phase space is 
the complex Grassmannian $\mathbb{G}^N_{M}=U(N)/[U(M)\times U(N-M)]$ for fractional values of $\nu=M$. The 4-component (spin-layer) CS on $\mathbb{G}^4_{2}$ for fractional values $\nu=2/\lambda$ of $\nu=2$ 
have been introduced in \cite{GrassCSBLQH,JPCMenredobicapa,JPA48} and 
recently extended to the $N$-component case $\mathbb{G}^N_{M}$ for filling factors $\nu=M/\lambda$ in \cite{APsigma}. In \cite{QPTBLQH} we have 
used these CS as variational states to study the classical limit and phase diagram of BLQH systems at $\nu=2/\lambda$. Here we are interested in the CS (phase-space or Bargmann) representation  
$\la Z|\psi\ra=\psi(Z)$ of quantum states $|\psi\ra$, the squared norm $Q_\psi(Z)=|\psi(Z)|^2$ being a positive quasi-probability distribution called the Husimi or $Q$-function. 
Both, Husimi and Wigner, phase-space quasi-probability distributions are useful to characterize phase-space properties of many quantum systems, specially in Quantum Optics  \cite{Leonhardt}, although 
Husimi  proves sometimes to be more convenient because, unlike Wigner, it is non-negative. It can also be measured by tomographic, spectroscopic, interferometric, etc, techniques, allowing a quantum state reconstruction. For example, 
one can visualize the time evolution of CS of light in a Kerr medium by measuring  $Q_{\psi}$ by cavity state tomography \cite{kirchmair2013}. 
Moreover, the zeros of the Husimi function have been used as an indicator of the
regular or chaotic behavior in quantum maps for a variety of atomic, molecular \cite{arranz2010,arranz2013}, condensed matter systems \cite{weinmann1999}, etc.
Information theoretic measures of $Q_\psi$ have also been considered as an indicator of metal-insulator  \cite{Aulbach} and topological-band insulator \cite{EPLWehrl} phase transitions, together with other QPTs  in 
Dicke, vibron, Lipkin-Meshkov-Glick (LMG), BEC and Bardeen-Cooper-Schrieffer (BCS) 
models \cite{romera2012,calixto2012,PRE2015,JSTAT2017,AnalCasta,pairons}, etc. In this article, we shall explore this phase-space tool to 
extract semi-classical information from the ground state of BLQH systems.

Using a CS representation, we shall obtain the Husimi function $Q_\psi(Z)$ of the ground state $\psi$ of a BLQH system at $\nu=2/\lambda$. This representation will allow us 
to visualize the structure of the ground state $\psi$ in phase-space in each of the three quantum phases (spin, canted and ppin). The Hamiltonian we shall use is an adaptation of the integer $\nu=2$ case \cite{HamEzawa} 
to the fractional $\nu=2/\lambda$ case \cite{QPTBLQH}. The localization of $\psi$ in phase-space can be quantified by the Husimi function second moment  $M_\psi=\int_{\mathbb{G}^4_{2}}Q^2_\psi(Z)d\mu(Z)$, where 
$d\mu(Z)$ denotes a proper measure on $\mathbb{G}^4_{2}$ (see later). Maximal localization (minimum volume/uncertainty) in phase space is attained when $\psi$ is itself a CS. This statement is in 
fact a conjecture that was proved for harmonic oscillator CS \cite{Wehrl79}  and recently  for the particular case of  $SU(2)$ spin-$s$ CS \cite{LiebAM}. Here we check the validity of this conjecture 
for Grassmannian $\mathbb{G}^4_{2}$ CS. In fact, we obtain that the ground state in spin and ppin phases is highly coherent (maximally localized), whereas it is more delocalized (higher uncertainty) in the canted phase, 
having the structure of a ``Schr\"odinger cat'', that is, a quantum superposition of two semi-classical states with negligible overlap [see later on eq. \eqref{parityadapt}].

The organization of the paper is as follows. In section \ref{secHaldane} we use the Haldane sphere picture for the (simpler) monolayer case to introduce some basic concepts like ``creation and annihilation operators 
of magnetic flux quanta'' and the coherent state (Bargmann-Fock) representation of quantum states, which will be essential to analyze the structure, semi-classical and localization properties of the ground state. 
In section \ref{sec2} we extend the spin-$s$ $U(2)$ symmetry to the isospin-$\lambda$ $U(4)$ symmetry, providing an oscillator realization of the $U(4)$ operators and the Landau-site Hilbert space for $\nu=2/\lambda$. We also 
review the isospin-$\lambda$ coherent states on $\mathbb{G}_2^4$, which are essential for the semiclassical ground state analysis of BLQH systems discussed in subsequent sections and to 
introduce the Husimi function and localization measures in phase space. Most of the construction has been already discussed in references \cite{GrassCSBLQH,JPCMenredobicapa,JPA48}; here we give a brief 
for the sake of self-containedness. In section \ref{sec3} we study 
the Landau-site Hamiltonian governing the BLQH system at $\nu=2/\lambda$, which is an adaptation of the one proposed in \cite{HamEzawa} for $\lambda=1$ to the fractional case (arbitrary odd $\lambda$); this Hamiltonian has already been discussed 
in \cite{QPTBLQH}. Using CS expectation values of the Hamiltonian, we perform a semiclassical analysis and a study of its quantum phases. In section \ref{sec4} we make a variational and exact (numerical diagonalization) ground state analysis 
and characterize the quantum phases (spin, ppin and canted) using localization measures in phase space. Variational results agree with numerical diagonalization calculations and provide analytical formulas for some physical quantities. 
The last section is left for conclusions and outlook.


\section{Haldane sphere U(2) picture}\label{secHaldane}

Firstly, we shall rephrase the simpler monolayer case in order to introduce the basic ingredients of the CS or Bargman picture. 
The technical innovation of Haldane \cite{Haldane} was to place the 2D electron gas on a spherical surface in a radial (monopole) 
magnetic field. Then, the total magnetic flux through the surface is an integer $2s$ times the flux quantum $\Phi_0=h/e$, as required 
by Dirac's monopole quantization. The Hilbert space of the lowest Landau level is spanned by polynomials in the spinor coordinates 
$u=\cos(\theta/2)\exp(\ic \phi/2)$ and $v=\sin(\theta/2)\exp(-\ic \phi/2)$ of total degree $2s$ ($\theta, \phi$ denote the 
polar and azimuthal angles on the sphere, respectively). Within this subspace, the electron may be represented by a spin $s$, the orientation of which 
indicates the point $(\theta,\phi)$ of the sphere about which the state is localized. Multiplication by $u$ and $v$ may also be represented 
as independent boson creation operators $a^\dag_\uparrow$ and  $a^\dag_\downarrow$ of magnetic flux quanta (flux quanta in the sequel) attached to the spin-up and 
spin-down electron respectively. In the same way, derivation by  $\partial/\partial_u$ and $\partial/\partial_v$ may also be represented 
as independent boson annihilation operators $a_\uparrow$ and  $a_\downarrow$ of flux quanta attached to the electron. This is 
related to the composite fermion picture \cite{Jainbook} of fractional QH effect, according to wich, bosonic flux quanta are attached to the electrons to form composite fermions.
The spin density 
operator $\vec{S}$ can be written in terms of these creation and annihilation operators of flux quanta as (the Jordan-Schwinger boson realization  for spin)
\be
S_+=a^\dag_\uparrow a_\downarrow,\; S_-=a^\dag_\downarrow a_\uparrow, \; S_3=(a^\dag_\uparrow a_\uparrow- a^\dag_\downarrow a_\downarrow)/2.\label{spinop}
\ee
This expression can be compactly written as 
\be {S}_\mu=\um \zeta^\dag\sigma_\mu \zeta, \;\; \mu=0,1,2,3,\label{spindensity}\ee
in terms of the two-component electron ``field''  $\zeta=\begin{pmatrix} a_\uparrow\\  a_\downarrow
\end{pmatrix}$ and its conjugate $\zeta^\dag=(a^\dag_\uparrow,  a^\dag_\downarrow)$, 
where $\sigma_\mu, \mu=1,2,3,$ denote the usual three Pauli matrices plus
$\sigma_0$ (the $2\times 2$ identity matrix). The four operators $S_\mu$ close the Lie algebra of $U(2)$. Actually, the extra operator  
$2 {S}_0=\zeta^\dag\zeta=a^\dag_\uparrow a_\uparrow +a^\dag_\downarrow a_\downarrow$ represents the total number $n_\uparrow+n_\downarrow=2s$ (twice the spin $s$) 
of flux quanta, which is conserved since $[S_0,\vec{S}]=0$. The spin third 
component $S_3$ measures the flux quanta imbalance between 
spin up and down, whereas $S_\pm=S_1\pm\ic S_2$ are tunneling (ladder) operators 
that transfer flux quanta from spin up to down and vice versa, creating spin coherence. 

The boson realization \eqref{spindensity} defines a unitary representation of the spin $U(2)$ 
operators $S_\mu$ on the Fock space expanded by the orthonormal basis states 
\be
|n_\uparrow\rangle\otimes|n_\downarrow\rangle=\frac{(a^\dag_\uparrow)^{n_\uparrow}(a^\dag_\downarrow)^{n_\downarrow}}{\sqrt{n_\uparrow!n_\downarrow!}}|0\ra_\mathrm{F}, \label{bosrepreu2}
\ee
where $|0\ra_\mathrm{F}$ denotes the Fock vacuum and $n_{\uparrow(\downarrow)}$ the number of flux quanta attached to spin up (down).  
The fact that $2 {S}_0$ is conserved  indicates that the representation
\eqref{spindensity} is reducible in Fock space. A $(2s+1)$-dimensional irreducible (Hilbert) subspace 
$\mathcal H_s(\mathbb S^2)$ carrying a unitary representation of $U(2)$ with spin $s$ is expanded 
by the $S_3$ eigenvectors
\be
|k\ra\equiv|s+k\rangle_\uparrow\otimes|s-k\rangle_\downarrow=
\frac{\varphi_k(a^\dag_\uparrow)}{\sqrt{\frac{(2s)!}{(s+k)!}}}
\frac{\varphi_{-k}(a^\dag_\downarrow)}{\sqrt{\frac{(2s)!}{(s-k)!}}}|0\ra_\mathrm{F},\label{basisinfocksu2}
\ee
with $k=-s,\dots,s$ the corresponding spin third component [flux quanta imbalance $(n_\uparrow-n_\downarrow)/2$] and $\varphi_k(z)=\binom{2s}{s+k}^{1/2}z^{s+k}$. We have made use of the monomials  
$\varphi_k(z)$ as a useful notation to generalize the Fock space representation 
\eqref{basisinfocksu2} of the spin-$s$ $U(2)$ states $|k\rangle$, to the 
isospin-$\lambda$  $U(4)$ states $|{}{}_{q_a,q_b}^{j,m}\ra$ in eq. \eqref{basisinfock2}. 
The monomials $\varphi_k(z)$ verify the closure relation 
\be \sum_{k=-s}^2\overline{\varphi_k(z')}\varphi_k(z)=K_{s}(\bar{z}',z),\label{bergmannu2}\ee
with $K_{s}(\bar{z}',z)=(1+\bar{z}'z)^{2s}$ the so-called Bergmann kernel for spin-$s$  [see \eqref{Bergmann} for its 
generalization to the bilayer case $U(4)$]. $\mathcal H_s(\mathbb S^2)$ is then the carrier space of the $(2s+1)$-dimensional 
 totally symmetric unitary irreducible representation that arises  in the Clebsch-Gordan decomposition of a tensor product of $2s$ two-dimensional (fundamental, elementary) representations of
$U(2)$; for example, in Young tableau notation:
\begin{equation}
   \overbrace{\begin{Young}  \cr \end{Young}  \otimes\dots\otimes \begin{Young}  \cr \end{Young}}^{2s}=
    \overbrace{\begin{Young}  & ... & \cr   \end{Young}}^{2s}\oplus\dots, \label{CGdecompu2}
\end{equation}
or $\overbrace{[1]\otimes\dots\otimes [1]}^{2s}=[2s]\oplus\dots$.  This Young tableau notation will be usefull when interpreting the 
bilayer case at fractional values of $\nu=2$ in a group theoretical context [see expression \eqref{CGdecomp}].

As already said, the operators $S_\pm$ create spin coherence, which can be described by spin-$s$ CS
\be
|z\rangle=\frac{e^{z {S}_+}|-s\rangle}{(1+|z|^2)^{s}}=
\frac{\sum_{k=-s}^s\varphi_k(z)|k\rangle}{(1+|z|^2)^{s}},\label{su2cs}
\ee
obtained as an exponential action of the rising operator ${S}_+$ on the lowest-weight state $|k=-s\rangle$ (namely, all 
flux quanta attached to spin down electron). The coherence strength $z=v/u=\tan(\theta/2)e^{-i\phi}$ is the quotient of the spinor coordinates 
$u$ and $v$ defined above and is related to the stereographic projection of a 
point $(\theta,\phi)$  of the sphere $\mathbb S^2=U(2)/U(1)^2$ onto the complex plane. In other words, the CS $|z\rangle$ 
is the rotation of the state $|k=-s\rangle$  about the axis $\vec{r}=(\sin\phi,-\cos\phi,0)$ in the $x-y$ plane by 
an angle $\theta$. Perhaps, a more familiar Fock-space representation of spin-$s$ 
CS [equivalent to \eqref{su2cs}]  is given as a two-mode Bose-Einstein condensate 
\be
|z\rangle=\frac{1}{\sqrt{(2s)!}}\left(\frac{a_\downarrow^\dag+za^\dag_\uparrow}{\sqrt{1+|z|^2}}\right)^{2s}|0\ra_\mathrm{F}.\label{su2BE}
\ee
In this context,  the polar angle $\theta$ is related to the population imbalance $s\cos\theta$ (the spin third 
component expectation value $\la z|S_3|z\ra$) between modes and the azimuthal angle $\phi$ is the relative phase (coherence). 
Both quantities can be experimentally determined in terms of matter-wave interference experiments (see e.g.\cite{Saba}). 

From the mathematical point of view, spin-$s$ CS are normalized (but not orthogonal), as can be seen from the CS overlap
\be\langle z'|z\rangle=K_{s}(\bar{z}',z)/[K_{s/2}(\bar{z}',z')K_{s/2}(\bar{z},z)],\label{su2overlap}\ee
written in terms of the Bergmann kernel \eqref{bergmannu2}. CS constitute an overcomplete set fulfilling 
the resolution of the identity $1=\int_{{\mathbb S^2}}|z\rangle\langle z|d\mu(z,\bar z)$, 
with $ d\mu(z,\bar z)=\frac{2s+1}{\pi}\sin\theta d\theta d\phi$ the solid angle.

Spin-$s$ CS have minimal uncertainty and therefore they are suitable to study the semi-classical, mean-field or thermodynamical limit 
of many spin systems, specially those undergoing a QPT. The semi-classical properties of a quantum spin state $|\psi\rangle$ are 
better described in a CS or Fock-Bargmann representation of any spin state $\psi\in\mathcal H_s(\mathbb S^2)$ defined as 
$\Psi(z)=K_{s/2}(\bar{z},z)\la \psi|z\ra$. For example, the basis states $|\psi\ra=|k\ra$ are represented by the monomials 
$\varphi_k(z)=\binom{2s}{s+k}^{1/2}z^{s+k}$ in \eqref{basisinfocksu2}, whereas a general 
spin state $|\psi\rangle=\sum_{k=-s}^sc_k|k\rangle$ is represented by a polynomial $\Psi(z)=\sum_{k=-s}^s\bar c_k\varphi_k({z})$ 
of degree $2s$ in ${z}$. Inside this CS picture, spin operators \eqref{spinop} are represented by differential operators 
\be
\mathcal S_+=-z^2 \frac{d}{dz} + 2sz,\; \mathcal S_-=\frac{d}{dz},\; \mathcal S_3=z \frac{d}{dz}-s,\label{difsu2}\ee
so that the following the identity $\mathcal S_i\Psi(z)=K_{s/2}(\bar{z},z)\la \psi|S_i|z\ra$ holds. In other words, 
$\mathcal S_i$ are the infinitesimal generators of M\"obius transformations $z'=(a z+b)/(cz+d)$ of $z$ under a $SU(2)$ group translation 
$U=\begin{pmatrix} a & b\\ c & d\end{pmatrix}$. This differential realization of the $SU(2)$ spin generators is useful 
for technical calculations like CS expectation values and matrix elements
\be
\la z'|S_i|z\ra=[K_{s/2}(\bar{z},z)K_{s/2}(\bar{z}',z')]^{-1}\mathcal S_i K_{s}(\bar{z}',z),\label{cssu2overlap}
\ee
which are reduced to simple derivatives of the Bergmann kernel. For example $\la z|S_3|z\ra=s(|z|^2-1)/(|z|^2+1)=-s\cos\theta$. We shall 
make extensive use of this relation when computing the energy surface (the CS Hamiltonian expectation value).

The probability density in this CS representation is the so-called Husimi quasiprobability distribution function 
$Q_\psi(z)=|\langle z|\psi\rangle|^2$.  Basically, $Q_\psi(z)$ is the probability to measure the spin third component 
$k=-s$ (all flux quanta attached to spin down electron) in $\psi$ in an orientation given by $(\theta,\phi)$. For a general 
spin state $|\psi\rangle=\sum_{k=-s}^sc_k|k\rangle$, the Husimi amplitude $\langle \psi|z\rangle$ is basically a polynomial in $z$ (except 
for a normalization factor) of degree $2s$, which can be determined by a finite number of measurements, thus allowing a state reconstruction. 

For normalized states $\langle\psi|\psi\rangle$, the resolution of the identity 
$1=\int_{{\mathbb S^2}}|z\rangle\langle z|d\mu(z,\bar z)$ indicates that the quasiprobability distribution $Q_\psi$ is normalized 
$\int_{{\mathbb S^2}}Q_\psi(z)d\mu(z,\bar z)=1$. The Husimi second moment 
\be M_\psi=\int_{{\mathbb S^2}}Q^2_\psi(z)d\mu(z,\bar z), \label{MQu2}\ee
also called ``inverse participation ratio'' (IPR), will be an important quantity for us. 
Broadly speaking, the IPR measures the spread of a state $|\psi\rangle$ over a basis  $\{|i\rangle\}_{i=1}^d$. 
Precisely, if $p_i$ is the 
probability of finding the (normalized) state $|\psi\rangle$ in $|i\rangle$, then the IPR is defined as $M_\psi=\sum_i p_i^2$. 
If $|\psi\rangle$ only ``participates'' of a single state $|i_0\rangle$, then 
$p_{i_0}=1$ and $M_\psi=1$ (large IPR), whereas if $|\psi\rangle$ equally participates on all of them (equally distributed), $p_{i}=1/{d}, \forall i$, then 
$M_\psi=1/d$ (small IPR). Therefore, the IPR is a measure of the localization of $|\psi\rangle$ in the corresponding basis. 
For our case, the Husimi second moment 
\eqref{MQu2} measures how close is $|\psi\ra$ to a coherent state $|Z\ra$. 
$M_\psi$ attains its maximum value $M_\mathrm{max}=1/2+1/(2+8s)$ (maximum localization) when $|\psi\ra$ 
is itself a (minimum uncertainty) CS (see \cite{LiebAM} for a proof). There are other localization measures of $\psi$, quantifying the area occupied 
by $Q_\psi$ in the sphere $\mathbb{S}^2$,  like 
the Wehrl entropy $W_\psi=\int_{{\mathbb S^2}}Q_\psi(z)\ln Q_\psi(z)d\mu(z,\bar z)$, but we shall use $M_\psi$ because 
it is easier to compute and provides similar qualitative information.

Localization measures defined in terms of the Husimi function have proved to be a good tool to analize QPTs in Hamiltonian systems writen in terms of $SU(2)$ collective generators $\vec{S}$ like, for example, 
Dicke, vibron, LMG, BEC and BCS models \cite{romera2012,calixto2012,PRE2015,JSTAT2017,AnalCasta,pairons}, etc. The Husimi function $Q_\psi$ provides essential information and, in particular, its zeros, 
which turn out to be related to pairing energies in LMG and BCS  pairing mean-field Hamiltonians. In the next sections, we use the Husimi function to extract information about the quantum phases that 
appear in BLQH systems at $\nu=2/\lambda$ for Hamiltonians written in terms of $U(4)$ collective operators $\vec{S}, \vec{P}$ and $\mathbf R$.


\section{Grassmannian U(4) picture for the bilayer case}\label{sec2}

\subsection{U(4) symmetry and bosonic flux quanta representation}

The bilayer case introduces a new degree of freedom (layer or pseudospin) to the electron and, therefore, BLQH systems underlie an isospin $U(4)$ symmetry. 
Inside the composite fermion picture exposed in the 
previous section,  bosonic magnetic flux quanta are attached to the electrons to form composite fermions in the fractional case. 
Let us denote by  $(a_l^\downarrow)^\dag$ [resp. $(b_l^\uparrow)^\dag$] creation operators of flux quanta  attached to the electron $l$ 
with spin down [resp. up] at layer $a$ [resp. $b$], and so on. For the case of filling factor $\nu=2$ (two electrons, $l=1, 2$, per Landau site) the  electron ``field'' 
$\zeta$ is now arranged as a four-component compound  $\zeta=(\zeta_1,\zeta_2)$ of two fermions.  The sixteen $U(4)$ density operators  are then 
written as bilinear products of creation and annihilation operators as [remember the expression \eqref{spindensity} for $U(2)$ spin operators]
\begin{equation}
{T}_{\mu\nu}=\tr({\zeta}^\dag\tau_{\mu\nu} \zeta), \;\; \zeta=\begin{pmatrix}
            \mathbf a\\ \mathbf b
           \end{pmatrix}=\begin{pmatrix}
\begin{matrix} a_1^\downarrow & a_2^\downarrow \\ a_1^\uparrow & a_2^\uparrow
\end{matrix}
\\ \begin{matrix} b_1^\uparrow & b_2^\uparrow\\ b_1^\downarrow & b_2^\downarrow
\end{matrix} \end{pmatrix},\label{calzeta}\end{equation}%
where the sixteen $4\times 4$ matrices $\tau_{\mu\nu}\equiv\sigma_\mu^{\mathrm{ppin}}\otimes\sigma_\nu^{\mathrm{spin}}, \, \mu,\nu=0,1,2,3$, 
denote the $U(4)$ generators in the four-dimensional fundamental representation [they are written as a tensor product of spin and pseudospin/layer (ppin for short) Pauli matrices]. 
In the BLQH literature (see e.g. \cite{EzawaBook}) it is customary to denote the total spin ${S}_k={T}_{0k}/2$ and
ppin ${P}_k={T}_{k0}/2$, together with the remaining 9 isospin  ${R}_{kl}={T}_{lk}/2$ operators for $k,l=1,2,3$.
A constraint in the Fock space of eight boson modes is imposed such that $\zeta^\dag \zeta=\lambda
I_{2}$, with $\lambda$ representing the number of flux quanta bound to each electron and $I_2$ the $2\times 2$ identity. 
In particular, the linear Casimir operator ${T}_{00}=\tr(\zeta^\dag \zeta)$, providing the total number of flux quanta,
is fixed to $n_{a}+n_b=\lambda+\lambda=2\lambda$, with $n_{a}=n_{a1}^{\uparrow}+n_{a1}^{\downarrow}+n_{a2}^{\uparrow}+n_{a2}^{\downarrow}$ 
the total number of flux quanta in layer $a$ (resp. in layer $b$). The quadratic Casimir operator is also fixed to
\begin{equation}
\vec{S}^2 + \vec{P}^2 + \mathbf{R}^2=\lambda(\lambda+4).\label{Casimir}
\end{equation}
We also identify the interlayer imbalance operator  ${P}_3$ (ppin third component), which measures the excess of flux quanta between
layers $a$ and $b$, that is $\frac{1}{2}(n_{a}-n_b)$. Therefore, the realization \eqref{calzeta} defines a unitary bosonic representation of the $U(4)$ matrix generators $\tau_{\mu\nu}$ in the Fock
space of eight modes with constrains.  The corresponding Hilbert space will be denoted by ${\cal H}_\lambda(\mathbb G_2^4)$, which generalizes  ${\cal H}_s(\mathbb S^2)$ for the monolayer case. 
The dimension of ${\cal H}_\lambda(\mathbb G_2^4)$ corresponds to the different ways to attach $2\lambda$ flux quanta to two identical electrons in two layers. 
The count of states is as follows. The first electron can occupy any of the four isospin states $|b\!\uparrow\rangle, |b\!\downarrow\rangle,
|a\!\uparrow\rangle$ and $|a\!\downarrow\rangle$ at one Landau site of the lowest Landau level. Therefore, there are $\tbinom{4+\lambda-1}{\lambda}$ ways of
distributing $\lambda$ quanta among these four states. Due to the Pauli exclusion principle, there are only three states
left for the second electron and $\tbinom{3+\lambda-1}{\lambda}$ ways of
distributing $\lambda$ flux quanta among these three states. However, some of the previous configurations must be identified
since both electrons are indistinguishable and $\lambda$ pairs of quanta adopt $\tbinom{2+\lambda-1}{\lambda}$
equivalent configurations. In total, there are
\begin{equation}
 d_\lambda=\frac{\binom{\lambda+3}{\lambda}\binom{\lambda+2}{\lambda}}{\binom{\lambda+1}{\lambda}}=\frac{1}{12}(\lambda+3)(\lambda+2)^2(\lambda+1)\nn
\end{equation}
ways to distribute $2\lambda$ flux quanta among two identical electrons in four states.  This is precisely the dimension of the rectangular Young tableau of shapes  
$[\lambda,\lambda]$ (2 rows of $\lambda$ boxes each) arising in the Clebsch-Gordan decomposition of a tensor product of $2\lambda$ four-dimensional (fundamental, elementary) representations of
$U(4)$
\begin{equation}
   \overbrace{\begin{Young}  \cr \end{Young}  \otimes\dots\otimes \begin{Young}  \cr \end{Young}}^{2\lambda}=
    \overbrace{\begin{Young}  & ... & \cr   & ... & \cr \end{Young}}^{\lambda}\oplus\dots, \label{CGdecomp}
\end{equation}
or $\overbrace{[1]\otimes\dots\otimes [1]}^{2\lambda}=[\lambda,\lambda]\oplus\dots$ This rectangular Young tableaux picture also arises in  $N$-component 
antiferromagnets \cite{AffleckNPB257,Sachdev,Arovas}.  
Note that quantum states associated to Young tableaux $[\lambda,\lambda]$ are antisymmetric (fermionic character) under the 
interchange of the two electrons (two rows) for $\lambda$ odd, whereas they are symmetric (bosonic character) for $\lambda$ even. Therefore, composite fermions require $\lambda$ odd. 

\subsection{Orthonormal basis and coherent states}
In Refs. \cite{GrassCSBLQH,JPCMenredobicapa} we have worked out an orthonormal basis 
\begin{equation}
B_\lambda(\mathbb G_2^4)=\left\{|{}{}_{q_a,q_b}^{j,m}\ra, \;\begin{matrix}
  2j, m\in\mathbb N,\\  q_a,q_b=-j,\dots,j \end{matrix}\right\}_{2j+m\leq\lambda},\label{basisvec}
\end{equation}
of the $d_\lambda$-dimensional carrier Hilbert space
${\cal H}_\lambda(\mathbb G_2^4)$, generalizing the spin $S_3$ eigenvectors $B_s(\mathbb S^2)=\{|k\ra, k=-s,\dots,s\}$ in eq. \eqref{basisinfocksu2}. The orthonormal basis vectors 
$|{}{}_{q_a,q_b}^{j,m}\ra$ are now indexed by four (half-)integer numbers subject to constraints. We shall provide here a brief summary with the basic expressions, in order to make the article more self-contained 
(more information can be found in references \cite{GrassCSBLQH,JPCMenredobicapa,JPA48,APsigma,QPTBLQH}).

Similar to \eqref{basisinfocksu2} for spin-$s$ states $|k\ra$, the general expression 
of these basis states $|{}{}_{q_a,q_b}^{j,m}\ra$ can be given by the action of creation operators $\mathbf{a}^\dag$ and $\mathbf{b}^\dag$ of flux quanta on the Fock vacuum $|0\ra_\mathrm{F}$ as 
\bea
|{}{}_{q_a,q_b}^{j,m}\ra&=&\frac{1}{\sqrt{2j+1}}\sum_{q=-j}^{j}(-1)^{q_a-q}\label{basisinfock2}\\
&\times&\frac{\varphi^{j,m}_{-q,-q_a}(\mathbf{a}^\dag)}{\sqrt{\frac{\lambda!(\lambda+1)!}{(\lambda-2j-m)!(\lambda+1-m)!}}}
\frac{\varphi^{j,\lambda-2j-m}_{q,q_b}(\mathbf{b}^\dag)}{\sqrt{\frac{\lambda!(\lambda+1)!}{m!(2j+m+1)!}}}
\;|0\ra_\mathrm{F},\nn
\eea
where 
\bea
\varphi_{q_a,q_b}^{j,m}(Z)&=&\sqrt{\frac{2j+1}{\lambda+1}\binom{\lambda+1}{2j+m+1}\binom{\lambda+1}{m}}\label{basisfunc}\\
&\times& \det(Z)^{m}\cD^{j}_{q_a,q_b}(Z),\; \begin{matrix}
2j+m\leq\lambda, \\ q_a,q_b=-j,\dots,j, \end{matrix}\nn\eea
are homogeneous polynomials of degree $2j+2m$ in four complex variables $z_{uv}\in\mathbb{C}$ arranged in a $2\times 2$ complex matrix $Z=\begin{pmatrix} z_{11} & z_{12}\\ z_{21} & z_{22}\end{pmatrix}$ 
(a point on the Grassmannian $\mathbb{G}^4_{2}$).  They generalize the monomials $\varphi_k(z)=\binom{2s}{s+k}^{1/2}z^{s+k}$ in \eqref{basisinfocksu2} for a point $z$ on the sphere $\mathbb S^2$. 
By $\cD^{j}_{q_a,q_b}(Z)$ we denote the usual Wigner $\cD$-matrix \cite{Louck3} with angular momentum $j$. They are homogeneous polynomials of 
degree $2j$ explicitly given by
\bea
&& \cD^{j}_{q_a,q_b}(Z)=\sqrt{\frac{(j+q_a)!(j-q_a)!}{(j+q_b)!(j-q_b)!}}
 \sum_{k=\max(0,q_a+q_b)}^{\min(j+q_a,j+q_b)}\label{Wignerf}\\ 
&& \binom{j+q_b}{k}\binom{j-q_b}{k-q_a-q_b}   z_{11}^k
z_{12}^{j+q_a-k}z_{21}^{j+q_b-k}z_{22}^{k-q_a-q_b}.\nn\eea
The closure relation \eqref{bergmannu2} now adopts the following form
\begin{equation}\sum^{\lambda}_{m=0}\!\!\sum_{j=0;\um}^{(\lambda-m)/2}\!\!\sum^{j}_{q_a,q_b=-j}\!\!
\overline{\varphi_{q_a,q_b}^{j,m}({Z'})}\varphi_{q_a,q_b}^{j,m}(Z)=K_\lambda(Z'^\dag, 
Z),\label{Bergmann}\end{equation}
with $K_\lambda(Z'^\dag,Z)=\det(\sigma_0+Z'^\dag Z)^\lambda$ the Bergmann kernel for $\mathbb G_2^4$.  
The orthonormal basis states \eqref{basisinfock2} are eigenstates of the following operators:
\bea
P_3|{}{}_{q_a,q_b}^{j,m}\ra&=&(2j+2m-\lambda)|{}{}_{q_a,q_b}^{j,m}\ra,\nn\\
(\vec{S}_a^2+\vec{S}_b^2)|{}{}_{q_a,q_b}^{j,m}\ra&=&2j(j+1)|{}{}_{q_a,q_b}^{j,m}\ra,\label{CCOC}\\
S_{\ell 3}|{}{}_{q_a,q_b}^{j,m}\ra&=&q_\ell|{}{}_{q_a,q_b}^{j,m}\ra,\; \ell=a, b,\nn
\eea
where we have defined angular momentum operators in layers $a$ and $b$ as  $S_{a k}=-\frac{1}{2}(S_k+R_{k3})$ and $S_{b k}=\frac{1}{2}(S_k-R_{k3})$, $k=1,2,3$, respectively, so that $\vec{S}_a^2+\vec{S}_b^2=\frac{1}{2}(\vec{S}^2+\vec{R}_3^2)$. 
Therefore, $j$ is a half-integer representing the total angular momentum of layers $a$ and $b$, whereas $q_a$ and $q_b$ are the corresponding third components. The integer $m$ is related to the interlayer imbalance 
population (ppin third component $P_3$) through $\frac{1}{2}(n_{a}-n_b)=(2j+2m-\lambda)$; thus, $m=\lambda, j=0$ means $n_a=2\lambda$ (i.e., all flux quanta occupying layer $a$), whereas  $m=0, j=0$ means $n_b=2\lambda$ (i.e., all flux quanta occupying layer $b$).  
The angular momentum third components $q_a, q_b$ measure the imbalance between spin up and down in each layer, more precisely, 
$q_a=\frac{1}{2}(n_{a1}^{\uparrow}-n_{a1}^{\downarrow}+n_{a2}^{\uparrow}-n_{a2}^{\downarrow})$ and similarly for $q_b$.

Analogously to the two-mode boson (flux quanta) condensate \eqref{su2BE}, 
Coherent states on $\mathbb{G}^4_{2}$ are defined as eight-mode boson condensates (see \cite{GrassCSBLQH}) 
\begin{equation}
|Z\ra=\frac{1}{\lambda!\sqrt{\lambda+1}}\left(\frac{\det(\check{\mathbf{b}}^\dag+
Z^t\check{\mathbf{a}}^\dag)}{\sqrt{\det(\sigma_0+Z^\dag Z)}}\right)^\lambda|0\ra_\mathrm{F},
\label{u4csfock}
\end{equation}
where $\check{\mathbf{a}}^\dag=\um\eta^{\mu\nu}\tr(\sigma_\mu\mathbf{a}^\dag)\sigma_\nu$ denotes the ``parity reversed'' 
$2\times 2$-matrix creation operator 
of $\mathbf{a}^\dag$ in layer $a$  (similar for layer $b$) [we are using Einstein summation convention with Minkowskian
metric $\eta^{\mu\nu}=\mathrm{diag}(1,-1,-1,-1)$ for notational convenience]. They can be expanded in the orthonormal basis \eqref{basisvec} as 
\begin{equation}
|Z\ra=\frac{\sum^{\lambda}_{m=0}\sum_{j=0;\um}^{(\lambda-m)/2}\sum^{j}_{q_a,q_b=-j}\varphi_{q_a,q_b}^{j,m}(Z)
|{}{}_{q_a,q_b}^{j,m}\ra}{\det(\sigma_0+Z^\dag Z)^{\lambda/2}},\label{u4cs}
\end{equation}
with coefficients $\varphi_{q_a,q_b}^{j,m}(Z)$ [compare to \eqref{su2cs} for the monolayer case]. Coherent states are normalized, $\la Z|Z\ra=1$, 
but they do not constitute an orthogonal set since they have a non-zero (in general) overlap given by
\begin{equation}
\la Z'|Z\ra=\frac{K_\lambda(Z'^\dag, Z)}{K_{\lambda/2}(Z'^\dag, Z')K_{\lambda/2}(Z^\dag, Z)},\label{u4csov}
\end{equation}
with $K_\lambda$ the Bergmann kernel in \eqref{Bergmann}.

Using  orthogonality properties of the homogeneous polynomials $\varphi_{q_a,q_b}^{j,m}(Z)$, a resolution of unity for isospin-$\lambda$ CS has been proved in \cite{GrassCSBLQH}, namely
$1=\int_{\mathbb G_2^4} |Z\ra\la Z|d\mu(Z,Z^\dag)$, with integration measure [compare with the $\mathbb S^2$ measure after \eqref{cssu2overlap}]
\begin{equation}
 d\mu(Z,Z^\dag)=\frac{12d_\lambda}{\pi^4}\frac{\prod_{u,v=1}^2 d\mathrm{Re}(z_{uv})d\mathrm{Im}(z_{uv})}{\det(\sigma_0+Z^\dag Z)^{4}}.\label{measureG2}
\end{equation}
Instead of the four complex coordinates $z_{uv}$, we shall use an alternative parametrization of $Z$ in terms of eight angles 
$\theta_{a,b},\vartheta_\pm\in[0,\pi)$ and $\phi_{a,b},\beta_{\pm}\in[0,2\pi)$, given by the following decomposition [the analogue of the stereographic projection $z=\tan(\theta/2)e^{\ic\phi}$ for $\mathbb S^2$]
\bea Z=V_a \begin{pmatrix}\xi_+  &0\\ 0& \xi_- \end{pmatrix} V_b^\dag,\; \xi_\pm=\tan\frac{\vartheta_\pm}{2}e^{\ic\beta_\pm},\nn\\ 
V_\ell=\begin{pmatrix}\cos\frac{\theta_\ell}{2}&-\sin\frac{\theta_\ell}{2}e^{\ic\phi_\ell}\\ \sin\frac{\theta_\ell}{2}e^{-\ic\phi_\ell} & \cos\frac{\theta_\ell}{2} 
\end{pmatrix},\, \ell=a,b, \label{paramang}
\eea
where $V_{a,b}$ represent rotations in layers $\ell=a,b$ (note their ``conjugated'' character). In this coordinate system, the integration measure \eqref{measureG2} can be alternatively written as
\begin{equation}
 d\mu(Z,Z^\dag)=\frac{3d_\lambda}{2^9\pi^4}
 (\cos\vartheta_+-\cos\vartheta_-)^2d\varOmega_+ d\varOmega_- d\Omega_a d\Omega_b,\label{measureG22}
\end{equation}
where $d\varOmega_\pm=\sin\vartheta_\pm d\vartheta_\pm d\beta_\pm$ and $d\Omega_{\ell}=\sin\theta_{\ell} d\theta_{\ell} d\phi_\ell$ ($\ell=a,b$) are  solid angle elements.

\subsection{Coherent state expectation values and localization in phase space}

As commented in section \ref{secHaldane}, for semi-classical considerations is more convenient a CS picture  than a Fock space realization of physical states. 
The Bargmann representation  of a general state $|\psi\ra\in{\cal H}_\lambda(\mathbb G_2^4)$ given by the overlap
$\psi(Z)\equiv \langle Z|\psi\rangle$ between $|\psi\ra$ and a general CS $|Z\ra$ like \eqref{u4cs}. For example, the Bargmann representation of the basis states 
$|\psi\ra=|{}{}_{q_a,q_b}^{j,m}\ra$ is given in terms of the homogeneous polynomials in \eqref{basisfunc} as $\psi(Z)=\varphi_{q_a,q_b}^{j,m}(Z)/K_{\lambda/2}(Z^\dag, Z)$. 
Inside this CS picture, the $U(4)$ isospin generators $\tau_{\mu\nu}$ are represented by differential operators $\mathcal{T}_{\mu\nu}$ [remember \eqref{difsu2} for the monolayer case]. 
They are the infinitesimal generators of M\"obius-like transformations  on $\mathbb G_2^4$ 
\be Z'=(A Z+B)(CZ+D)^{-1}, \;U=\left(\ba{cc} A& B\\ C &D\ea\right),\label{difsu4}\ee
under $U\in U(4)$ group translations.  For example, it is 
easy to to see that the differential realization of the imbalance ppin generator $\tau_{k0}/2$ is given by 
$\mathcal{P}_3=z^\mu\partial_\mu-\lambda$, where $z^\mu=\tr(Z\sigma_\mu)/2, \mu=0,1,2,3$. We are using Einstein summation convention and denoting $\partial_\mu={\partial}/{\partial z^\mu}$ and 
$z_\nu=\eta_{\nu\mu}z^\mu$, with $\eta_{\nu\mu}=\mathrm{diag}(1,-1,-1,-1)$ the Minkowskian metric]. In addition, 
spin $\mathcal{S}_k$ and $\mathcal{R}_{k3}$ are written in terms of the Lorentz-like generators 
$\mathcal{M}_{\mu\nu}=z_\mu \partial_\nu-z_\nu \partial_\mu$ as $\mathcal{S}_i=\frac{\ic}{2}\epsilon^{ikl}\mathcal{M}_{kl}$ and  
$\mathcal{R}_{k3}=\mathcal{M}_{k0}$, respectively, where $\epsilon^{ikl}$ is the totally antisymmetric tensor.  The explicit expression of the remainder  $U(4)$ differential 
operators $\mathcal{T}_{\mu\nu}$ can be seen in \cite{GrassCSBLQH}. Some readers can wonder where this relativistic notation comes from. It is motivated by the fact that $U(4)$ is the compact counterpart of the conformal 
group $U(2,2)$, which contains the Poincar\'e group of special relativity; however, the compacity of $U(4)$ introduces some notational differences with respect to $U(2,2)$ like, for example, 
the ``parity reversal'' operation in some expressions like \eqref{u4csfock} [the reader can consult \cite{confosci} for the non-compact $U(2,2)$ case]. We just use relativistic notation for convenience.

As we already said in \eqref{cssu2overlap}  for the monolayer case, with this differential realization, the (cumbersome) computation of expectation values of operators in a coherent state (usually related to 
order parameters in the mean-field approximation) is reduced to the (easy) calculation of derivatives of the Bergmann kernel \eqref{Bergmann} as: 
\begin{equation}\la Z|T_{\mu\nu}|Z\ra=K_{\lambda}^{-1}(Z,Z^\dag)\mathcal{T}_{\mu\nu}K_{\lambda}(Z,Z^\dag).\end{equation}
We shall use this simple formula to compute the energy surface (the CS expectation value  $\la Z|H|Z\ra$ of the Hamiltonian $H$). For example, in terms of  ${M}_{\mu\nu}=2\ic\lambda\frac{
z_\mu \bar z^\nu-z_\nu \bar z^\mu}{\det(\sigma_0+Z^\dag Z)}$, the CS expectation values of spin and ppin operators turns out to be
\bea
&&\la S_1\ra= M_{23}, \, \la S_2\ra= M_{31}, \,\la S_3\ra= M_{12},\nn\\
&&\la {R}_{k3}\ra=\ic M_{0k},\; \la \vec{S}\ra^2+\la \vec{R}_3\ra^2=M_{\mu\nu}M^{\mu\nu}/2,\label{expectv}\\
&&\la {P}_1\ra={\lambda}\,{\mathrm{Re}[\tr(Z)(1+\det(Z^\dag)]}/{\det(\sigma_0+ZZ^\dag)}, \nn\\ && \la P_3\ra= \lambda({\det(Z^\dag Z)-1})/{\det(\sigma_0+Z^\dag Z)},\nn
\eea
where $\mathrm{Re}$ denotes the real part [$\la {P}_2\ra$ corresponds to the imaginary part] and $\ic$ is the imaginary unit. Note that the following identity for the magnitude of the $SU(4)$ 
isospin is automatically fulfilled for coherent state expectation values:
\begin{equation}
 \la \vec{S}\ra^2+\la \vec{P}\ra^2+\la \mathbf{R}\ra^2=\lambda^2.\label{isomag}
\end{equation}
For $\lambda=1$ it coincides with the variational ground state condition provided in \cite{HamEzawa}. For BLQH systems at $\nu=2/\lambda$ we have seen in \cite{QPTBLQH} that the spin and ppin 
phases are characterized by maximum values of $\la\vec{S}\ra^2=\lambda^2$ and $\la\vec{P}\ra^2=\lambda^2$, respectively.

The Husimi function $Q_\psi(Z)$ of a given state $|\psi\ra$ is the CS expectation value $Q_\psi(Z)=\la Z|\rho|Z\ra$ of the corresponding density 
matrix $\rho=|\psi\ra\la \psi|$ (this definition can be directly extended to mixed states). $Q_\psi(Z)$ provides the probability of finding $|\psi\ra$ in a coherent state $|Z\ra$. For example, 
for $|\psi\ra=|{}{}_{q_a,q_b}^{j,m}\ra$ the Husimi function follows a multivariate distribution function \cite{JPA48}. Taking into account the CS closure relation 
$1=\int_{\mathbb G_2^4} |Z\ra\la Z|d\mu(Z,Z^\dag)$ with integration measure \eqref{measureG2}, we see that $Q_\psi$ is normalized according to $\int_{\mathbb{G}_2}Q_\psi(Z)d\mu(Z,Z^\dag)=1$ for 
unit norm states $\la\psi|\psi\ra=1$.  The  Husimi second moment is defined as [compare with the monolayer case in \eqref{MQu2}]
\be
M_\psi=\int_{\mathbb{G}_2}Q_\psi^2(Z)d\mu(Z,Z^\dag).\label{MQ}\ee
As for the monolayer case, we shall conjecture that $M_\psi$ attains its maximum value (maximum localization) when $|\psi\ra$ is itself a CS. 
This conjecture has been proved for harmonic oscillator CS \cite{Wehrl79} and spin-$s$ or $SU(2)$ CS \cite{LiebAM}. 
We have calculated this maximum value for each $\lambda$ in \cite{JPA48} and it turns out to be:
\be
 M_{\mathrm{max}}(\lambda)=\frac{1}{16}-\frac{1/2}{1+\lambda}
 +\frac{45/32}{1+2\lambda}+\frac{3/32}{3+2\lambda},\label{MQmax}
 \ee
which tends to $M_{\mathrm{max}}(\infty)= 1/16$ for high isospin $\lambda$ values.  
Here we shall see that the ground state of a BLQH system attains this maximum value in spin a ppin phases, thus indicating that it is maximally localized in these phases (see next section). 

\section{Model Hamiltonian and quantum phases}\label{sec3}

In Ref. \cite{QPTBLQH} we have analyzed the ground state structure of BLQH at $\nu=2/\lambda$. The Hamiltonian we used is an adaptation of 
the Landau-site Hamiltonian for $\nu=2$ considered in \cite{HamEzawa}
\begin{equation}
H=H_\mathrm{C}+H_\mathrm{ZpZ}.\label{Ham}
\end{equation}
which consists of Coulomb and a combination of Zeeman and pseudo-Zeeman interactions. Discarding $U(4)$-invariant terms, the Coulomb part
\begin{equation}
H_\mathrm{C}=4\varepsilon_{\mathrm{D}}^-P_3^2-2\varepsilon_{\mathrm{X}}^-(\vec{S}^2+\vec{R}^2_3+P_3^2),
\end{equation}
is a sum of the naive capacitance ($\varepsilon_{\mathrm{D}}^-$) and the exchange ($\varepsilon_{\mathrm{X}}^-$) interactions. The  exchange and capacitance energy gaps are 
given in terms of the interlayer distance $\delta$ by
\begin{equation}
\varepsilon_{\mathrm{X}}^{\pm}=\frac{1}{4}\sqrt{\frac{\pi}{2}}\left(1\pm e^{(\delta/\ell_B)^2/2}\mathrm{erfc}\left(\frac{\delta}{\sqrt{2}\ell_B}\right)\right)\mathcal{E}_C,
\end{equation}
and $\varepsilon_{\mathrm{D}}^-=\frac{\delta}{4\ell_B}\mathcal{E}_C$, where $\mathcal{E}_C=e^2/(4\pi\epsilon \ell_B)$ is the Coulomb energy unit and $\ell_B=\sqrt{\hbar c/(eB)}$ the magnetic length. 
In the following we shall simply put $\varepsilon_{\mathrm{X}}^-=\varepsilon_{\mathrm{X}}$ and $\varepsilon_{\mathrm{D}}^-=\varepsilon_{\mathrm{D}}$ as no confusion will arise. 
We shall usually choose $\delta=\ell_B$, which gives $\varepsilon_{\mathrm{X}}\simeq 0.15$ in Coulomb units (we shall use Coulomb units throughout the article unless otherwise stated). 
The (pseudo) Zeeman part
\begin{equation}
H_\mathrm{ZpZ}= -\Delta_\mathrm{Z} S_3 - \Delta_\mathrm{t} P_1 - \Delta_\mathrm{b} P_3 \label{HamZpZ}
\end{equation}
is comprised of: Zeeman ($\Delta_\mathrm{Z}$),  interlayer tunneling  ($\Delta_\mathrm{t}$, also denoted by $\Delta_\mathrm{SAS}$ in the 
literature \cite{EzawaBook}) and bias ($\Delta_\mathrm{b}$) gaps. The bias term creates an imbalanced configuration between layers.

For $\nu=2/\lambda$ ($N=2\lambda$ flux quanta), Coulomb (two-body) interactions must be renormalized by the number of boson pairs $N(N-1)$, whereas one-body interactions must be renormalized by $N$, 
in order to make the energy an intensive quantity. Therefore, the Hamiltonian proposed for arbitrary $\lambda$ is an adaptation of  \eqref{Ham} of the form
\begin{equation}
H_\lambda=\frac{H_\mathrm{C}}{N(N-1)}+\frac{H_\mathrm{ZpZ}}{N}, \quad N=2\lambda.\label{Hamlambda}
\end{equation}
To study the semiclassical limit, we now replace the operators $P_j, S_j$ and $R_{ij}$ 
by their expectation values \eqref{expectv} in a isospin-$\lambda$ coherent state $|Z\ra$. 

A minimization process of the ground state energy surface $\la Z|H_\lambda|Z\ra$ reveals the existence of 
three quantum phases: spin, canted and ppin, which are characterized by maximum and minimum values of the squared spin $\la \vec{S}\ra^2$ and 
squared ppin $\la \vec{P}\ra^2$ CS expectation values (order parameters). For the sake of simplicity, let us restrict ourselves, for this semiclassical analysis, to the balanced case 
(i.e. we discard therms proportional to $\varepsilon_{\mathrm{D}}$ and $\Delta_\mathrm{b}$). Using the parametrization \eqref{paramang} of $Z$, we  found in \cite{QPTBLQH} 
the common relations
\begin{equation}\beta_+ = \beta_- = 0,~\vartheta_+ + \vartheta_-=\pi,~\theta_a + \theta_b = \pi,~\phi_a=\phi_b.\label{restrictangle} \end{equation}
in all phases. This leaves only two free parameters, for instance, $\vartheta_+$ and $\theta_b$. In the spin and ppin phases we have
\begin{equation}
  \mathrm{Spin:}\; \vartheta_+^s=0=\theta_a^s,\;  \mathrm{Ppin:}\;\vartheta_+^p=-\pi/2=\theta_a^p,\label{anglespinppin}
\end{equation}
respectively. In the canted phase we get the more involved expression
\bea
\tan\vartheta_+^c &=&  \pm\sqrt{\frac{(\DS^2-\DZ^2)^2-(4\DZ \varepsilon_{\mathrm{X}}({\lambda}))^2}
{-(\DS^2-\DZ^2)^2+(4\DS \varepsilon_{\mathrm{X}}({\lambda}))^2}}, \label{thetacanted}\\
\tan\theta_b^c &=& \mp\frac{\DS}{\DZ} \sqrt{\frac{(\DS^2-\DZ^2)^2-(4\DZ \varepsilon_{\mathrm{X}}({\lambda}))^2}
{-(\DS^2-\DZ^2)^2+(4\DS \varepsilon_{\mathrm{X}}({\lambda}))^2}},\nn
\eea
where we have defined $\varepsilon_{\mathrm{X}}({\lambda})=\lambda\varepsilon_{\mathrm{X}}/(2\lambda-1)$. The phase transition points (spin-canted and canted-ppin) depend 
on $\lambda$ and are located at 
\begin{eqnarray}
\DS^\mathrm{sc}(\lambda)&=&\sqrt{\DZ^2+4\varepsilon_{\mathrm{X}}(\lambda) \DZ},\label{ptp}\\ 
\DS^\mathrm{cp}(\lambda)&=&2\varepsilon_{\mathrm{X}}(\lambda)+\sqrt{\DZ^2+4\varepsilon_{\mathrm{X}}^2(\lambda)}.\nn
\end{eqnarray}
For $\DS<\DS^\mathrm{sc}(\lambda)$ the BLQH system at $\nu=2/\lambda$ is in the spin phase, 
for $\DS^\mathrm{sc}(\lambda)\leq \DS\leq \DS^\mathrm{cp}(\lambda)$ it is in the canted phase and for $\DS>\DS^\mathrm{cp}(\lambda)$ it is in the ppin phase [see \cite{QPTBLQH} for more details]. 
Note that we have two different solutions of $(\vartheta_+^c,\theta_b^c)$ in the canted phase, given by the 
signs $(+,-)$ and $(-,+)$ in equation \eqref{thetacanted}, leading to the same minimum energy $\la Z^c_\pm|H_\lambda|Z^c_\pm\ra$, with $Z^c_\pm=Z(\theta_{a,b},\phi_{a,b},\vartheta_\pm,\beta_\pm)|_\pm^c$ 
the corresponding stationary point in the Grassmannian $\mathbb{G}^4_{2}$ for any of the two solutions  $(+)=(+,-)$ and $(-)=(-,+)$ together with the common restrictions \eqref{restrictangle}. 
Even though both coherent states $|Z^c_+\ra$ and $|Z^c_-\ra$ give the same CS energy expectation value $\la Z^c_+|H_\lambda|Z^c_+\ra=\la Z^c_-|H_\lambda|Z^c_-\ra$ (see \cite{QPTBLQH}), they are 
distinct; in fact, they are almost orthogonal $\la Z^c_+|Z^c_-\ra\simeq 0$ in the canted phase. This indicates that the ground state is degenerated and there is a broken symmetry in the thermodynamic limit. 
Let us study the ground state structure in the phase-space (Bargmann) picture of section \ref{sec2}.

\section{Ground state analysis in phase-space and localization measures}\label{sec4}

\subsection{Variational results}

\begin{figure}[h]
\begin{center}
\includegraphics[width=4.2cm]{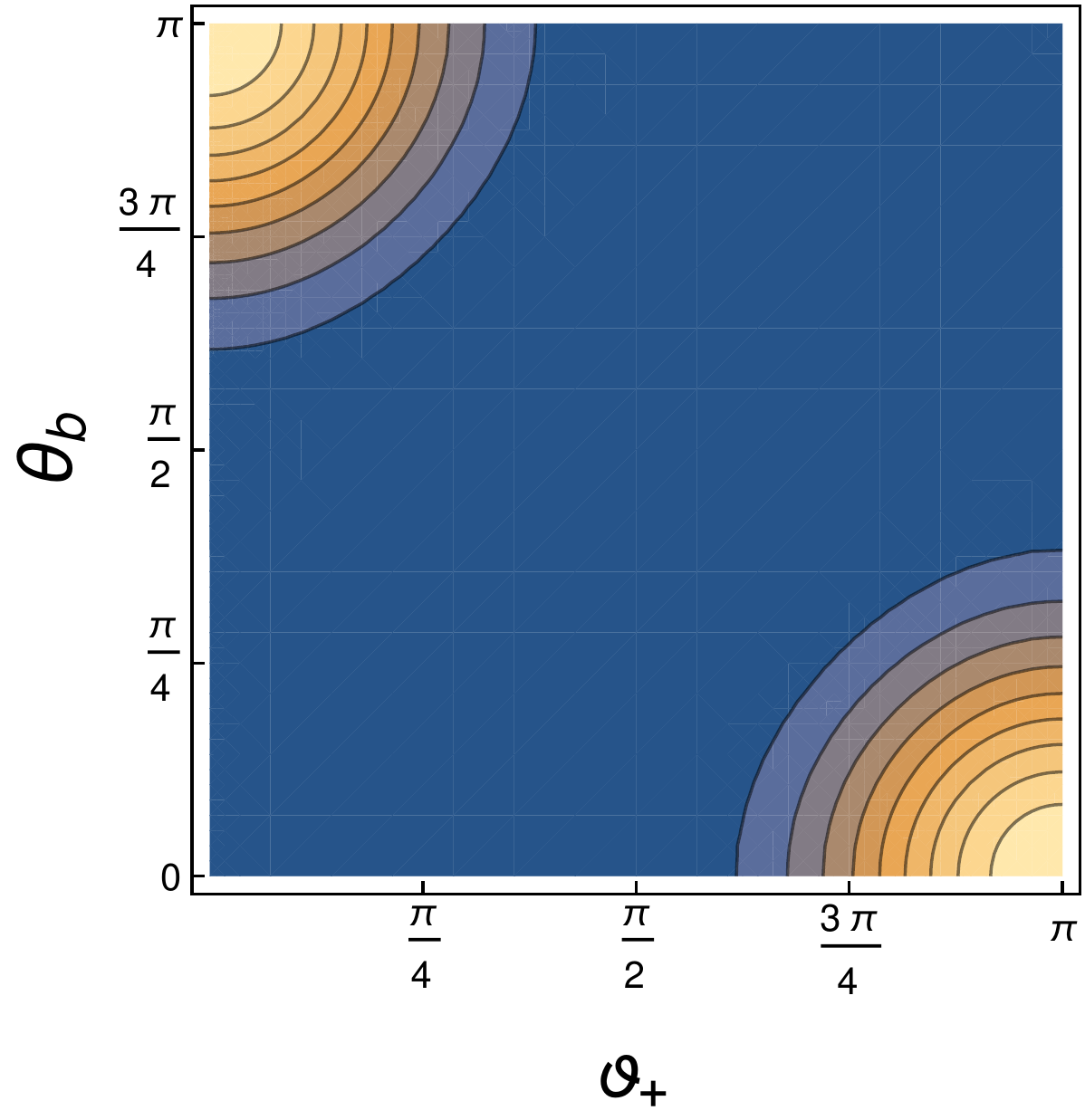} \includegraphics[width=4.2cm]{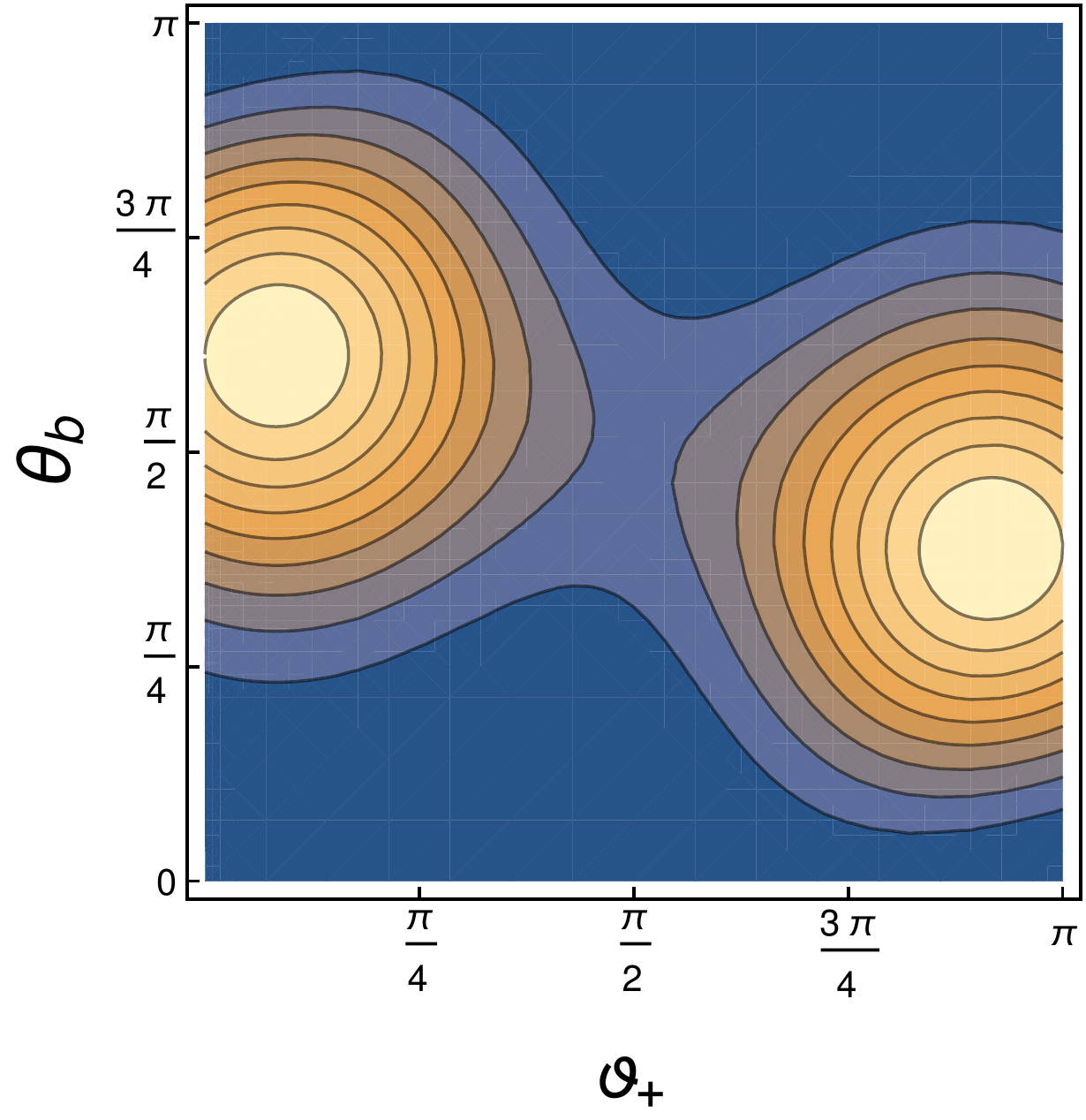}\\ 
\includegraphics[width=4.2cm]{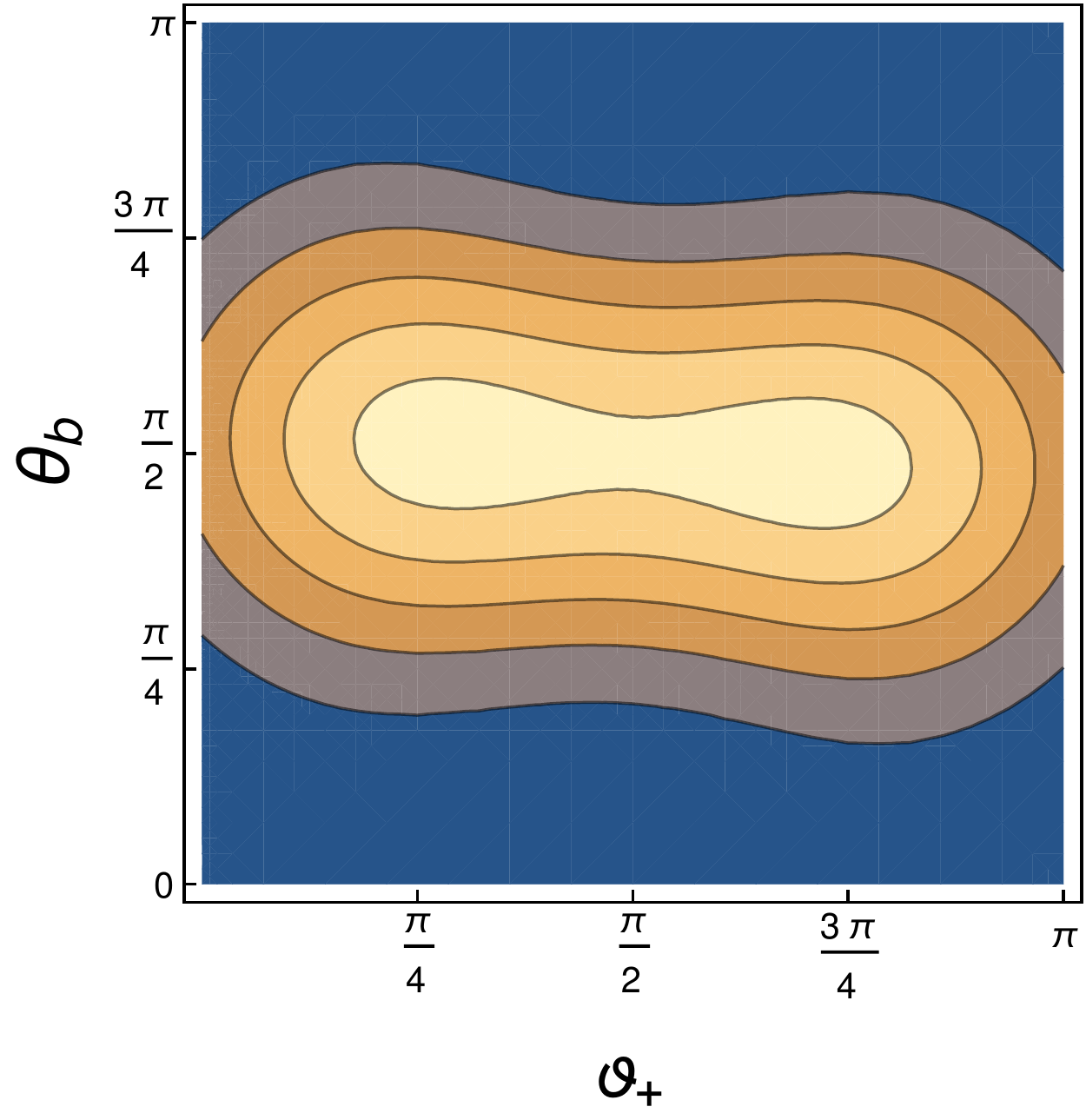} \includegraphics[width=4.2cm]{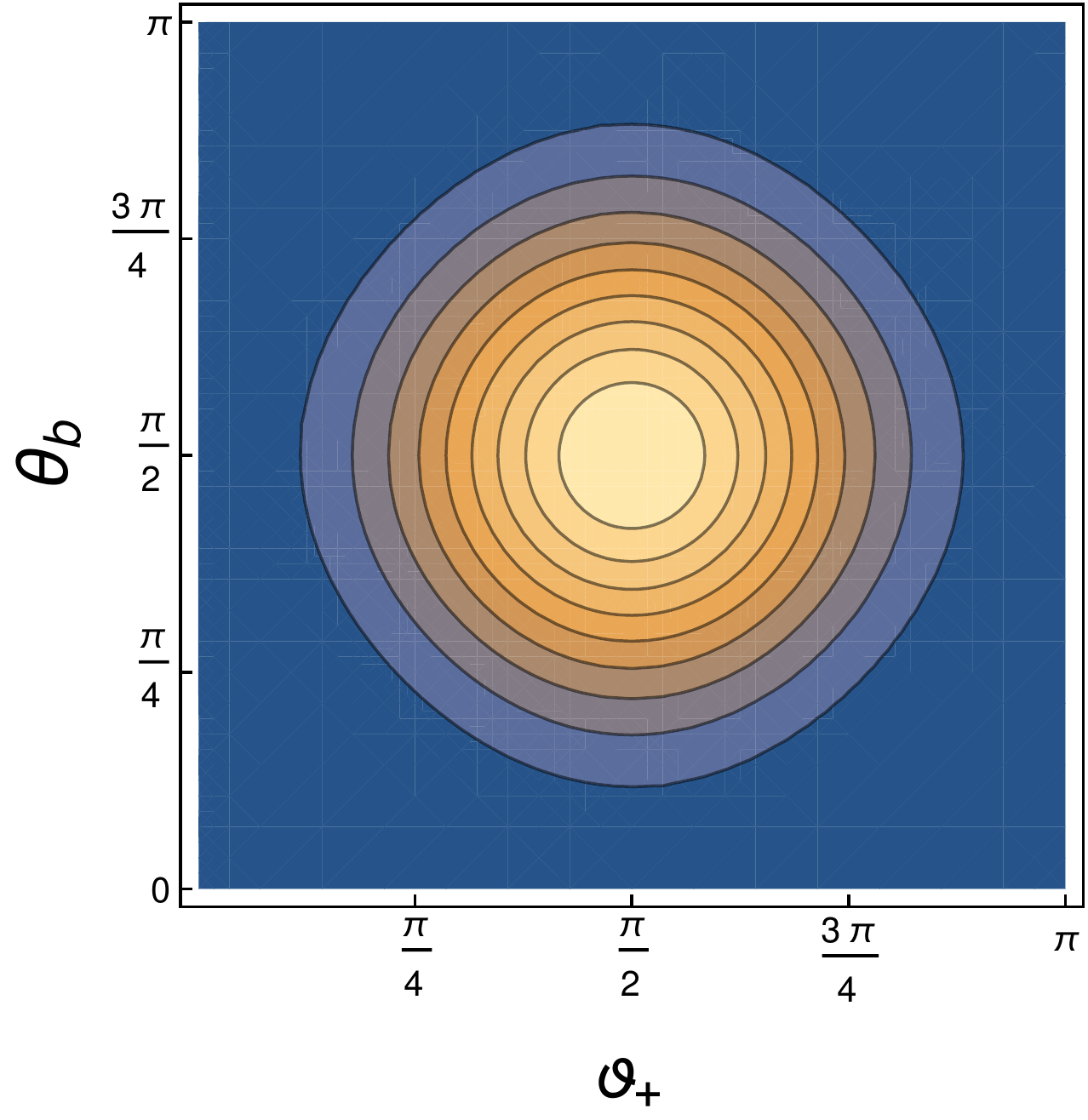}
\end{center}
\caption{Contour plot of the Husimi function \eqref{Husiadapt} of the variational ground state 
in the plane  $(\vartheta_+,\theta_b)$ of the phase-space $\mathbb{G}^4_2$ for $\lambda=3$, Zeeman $\DZ=0.01$, layer distance $\delta=\ell_B$ and four values of tunneling gap $\DS$.
The top-left panel corresponds to the spin phase ($\DS=0.01$), the top-right and bottom-left panels correspond to the canted phase 
($\DS=0.1$ and $\DS=0.2$, respectively) and the bottom-right panel corresponds to the canted phase ($\DS=0.5$). Lighter zones correspond to higher values of the Husimi function, that is, to higher 
probability for the ground state to be coherent.}\label{contourvar}
\end{figure}

We start with the analysis of the variational ground state. Let us denote collectively by $Z^0_+$ and $Z^0_-$ the two sets of stationary points in any of the three (spin, canted and ppin) quantum phases 
(note that $Z^0_+=Z^0_-$ in the spin and ppin phases). A good variational approximation to the true ground state is achieved by taking the normalized symmetric combination
\begin{equation}
 |Z^0_\mathrm{sym}\ra=\frac{|Z^0_+\ra+|Z^0_-\ra}{\sqrt{2(1+\mathrm{Re}(\la Z^0_+|Z^0_-\ra})}.\label{parityadapt}
\end{equation}
This is a quantum superposition of two coherent (semiclassical) states. 
Using the general expression of the CS overlap \eqref{u4csov} and the Bergmann kernel $K_\lambda(Z'^\dag,Z)=[1+\tr(Z'^\dag Z)+\det(Z'^\dag Z)]^\lambda$, we can easily compute the corresponding Husimi function 
\begin{equation}
Q^0_\mathrm{sym}(Z)=|\la Z^0_\mathrm{sym}|Z\ra|^2=\frac{|\la Z^0_+|Z\ra+\la Z^0_-|Z\ra|^2}{{2(1+\mathrm{Re}(\la Z^0_+|Z^0_-\ra})}.\label{Husiadapt}
\end{equation}
We shall restrict, for the sake of simplicity, to the plane  $(\vartheta_+,\theta_b)$ of the 8-dimensional Grassmannian 
phase-space $\mathbb{G}^4_2$ with constraints \eqref{restrictangle}, where non-trivial angle values \eqref{thetacanted} are found in the canted phase. 
In the plane  $(\vartheta_+,\theta_b)$, the Husimi function adopts a quite simple form given by
\begin{eqnarray}
Q^0_\mathrm{sym}(\vartheta_+,\theta_b)=\frac{(\cos(\vartheta_+-\vartheta_+^0)+\cos(\theta_b-\theta_b^0))^{2\lambda}}{2^{2\lambda}}, 
\end{eqnarray}
where $(\vartheta_+^0,\theta_b^0)$ must be replaced by \eqref{anglespinppin} and \eqref{thetacanted} in the spin, ppin and canted phases, respectively. 
In Figure \ref{contourvar} we represent a contour plot of $Q^0_\mathrm{sym}(\vartheta_+,\theta_b)$ in the three phases. We see that the variational state is localized around $(\vartheta_+^s,\theta_b^s)=(0,0)$ 
[or equivalently $(\vartheta_+^s,\theta_b^s)=(\pi,\pi)$] in the spin phase, 
and around $(\vartheta_+^p,\theta_b^s)=(\pi/2,\pi/2)$ in the ppin phase. In both, spin and ppin, phases we have $|Z^0_+\ra=|Z^0_-\ra$ and therefore $|Z^0_\mathrm{sym}\ra$ is coherent. 
In the canted phase, the variational ground state splits into two different packets, $|Z^c_+\ra\not=|Z^c_-\ra$,  localized around the 
two stationary solutions \eqref{thetacanted}. Both packets have negligible overlap  $\la Z^c_+|Z^c_-\ra\simeq 0$ and recombine in the spin and ppin regions. This kind of quantum superpositions of two 
semiclassical states with negligible overlap is sometimes referred to as a ``Schr\"odinger cat'' state in the literature, and they have many interesting physical properties in quantum information processing. 

\begin{figure}[h]
\begin{center}
\includegraphics[width=\columnwidth]{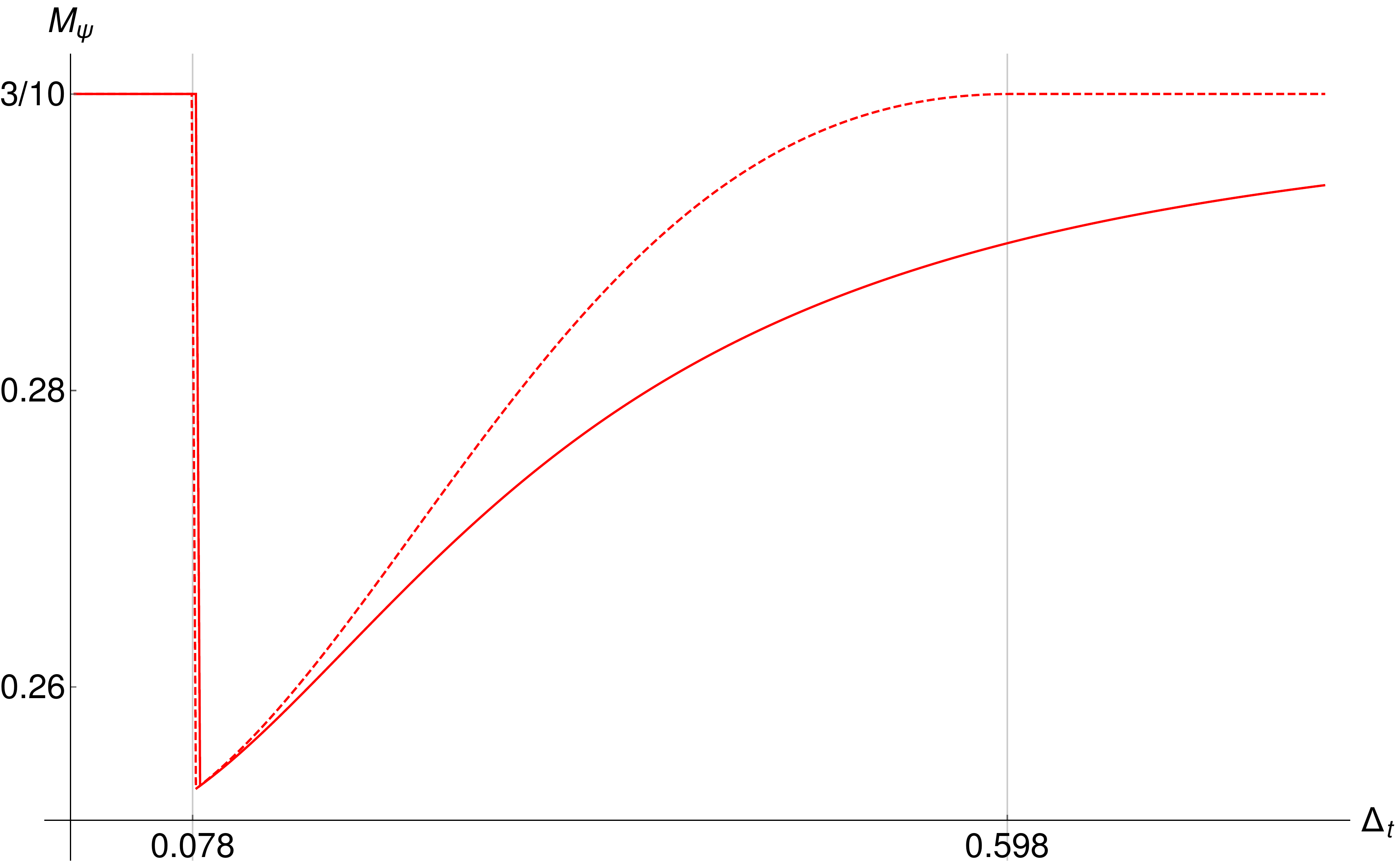}
\includegraphics[width=\columnwidth]{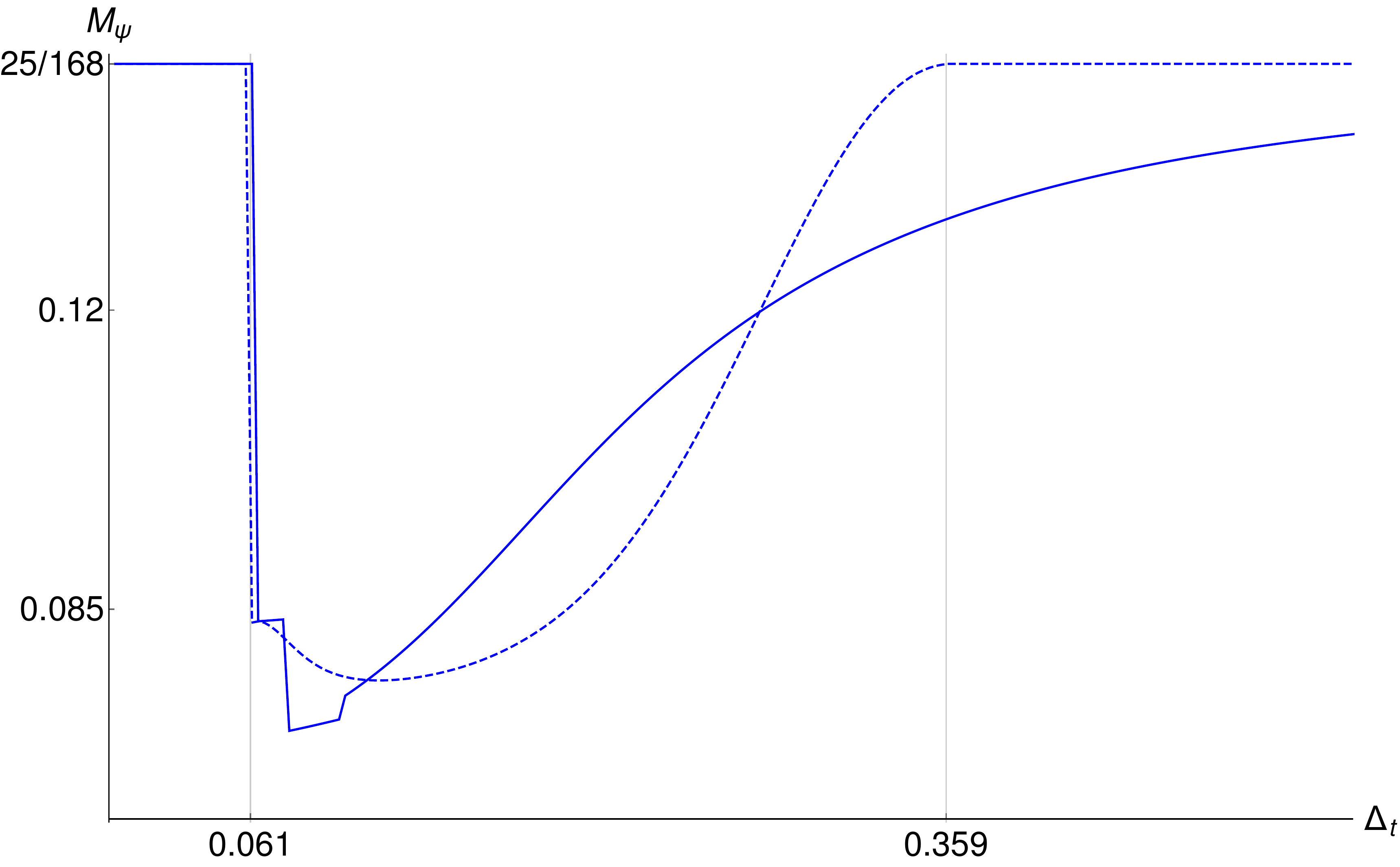}
\end{center}
\caption{Second moment $M_\psi$ of the Husimi function $Q_\psi$ of the variational (dashed) and exact (solid) ground states $\psi$ as a function of the tunneling $\DS$ for Zeeman $\DZ=0.01$, interlayer distance $\delta=\ell_B$ and $\lambda=1$ 
(top red) and $\lambda=3$ (bottom blue). Maximum moments values \eqref{MQmax} for $\lambda=1$ and $\lambda=3$ are $3/10$ and $25/168$, respectively. 
Spin-canted, $\Delta^{\mathrm{sc}}_\mathrm{t}(1)=0.078$ and  $\Delta^{\mathrm{sc}}_\mathrm{t}(3)=0.061$, and canted-ppin, $\Delta^{\mathrm{cp}}_\mathrm{t}(1)=0.598$ and  $\Delta^{\mathrm{cp}}_\mathrm{t}(3)=0.359$, 
phase-transition points \eqref{ptp} are marked by vertical dotted grid lines.}\label{MomentDt}
\end{figure}

This delocalization of $|Z^0_\mathrm{sym}\ra$ in phase-space inside the canted phase is captured by the Husimi function second moment \eqref{MQ}. Indeed, in figure \ref{MomentDt}  we represent 
the localization of the  variational and exact (see next section) ground state in phase-space measured by the Husimi second moment as a function of the tunneling $\DS$ (we fix $\DZ=0.01$ and $\delta=\ell_B$). 
We compare the two cases: $\lambda=1$  and $\lambda=3$. In the spin and ppin phases we have maximum localization [maximum moment \eqref{MQmax}], giving $M_{\mathrm{max}}(1)=3/10$ and $M_{\mathrm{max}}(3)=25/168$,  
since the ground state is a (minimal uncertainty) coherent state [note that, in the exact case, the maximum moment value is only attained asymptotically in the ppin phase]. In the transition from the spin to the canted phase 
we observe a sudden delocalization (a drop of the Husimi second moment) of the ground state wave function in phase-space. Therefore, the canted region is characterized  for having a much more delocalized 
ground state than in the spin and ppin regions. Thus, we conclude that the Husimi second moment serves as an order parameter characterizing the three phases and the phase transition points.

\begin{figure}[ht]
\begin{center}
\includegraphics[width=4.2cm]{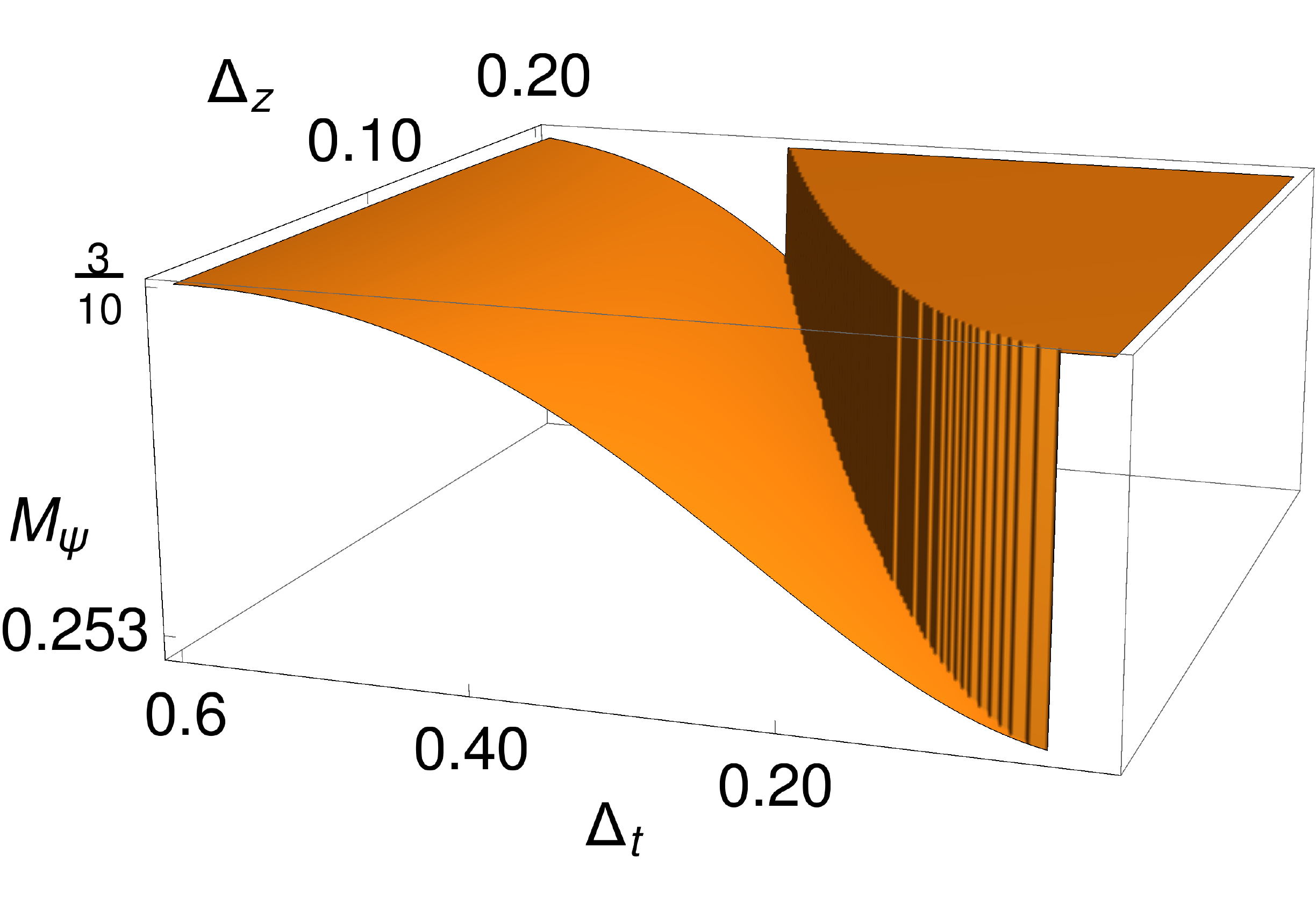} \includegraphics[width=4.2cm]{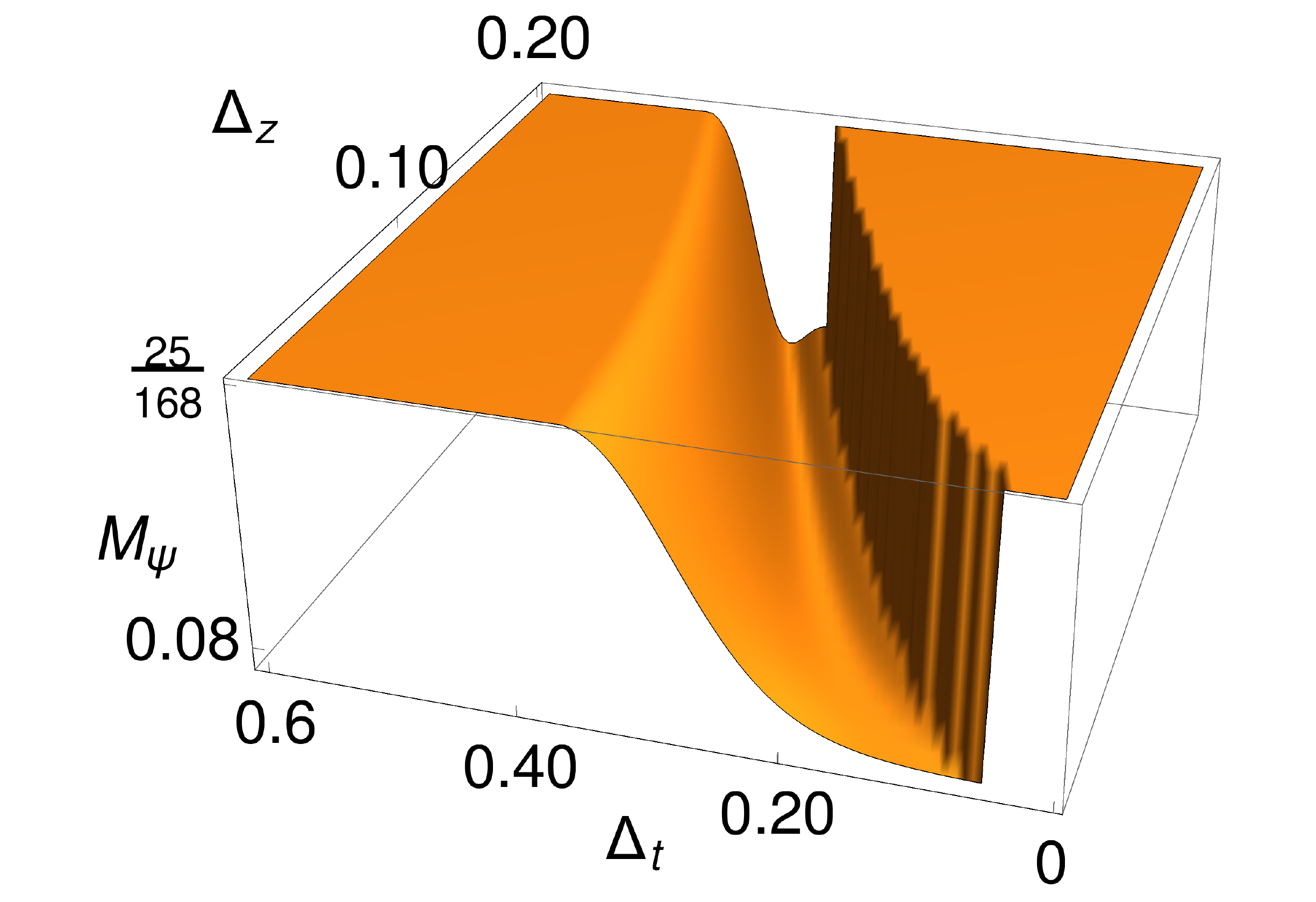}\\
\includegraphics[width=4.2cm]{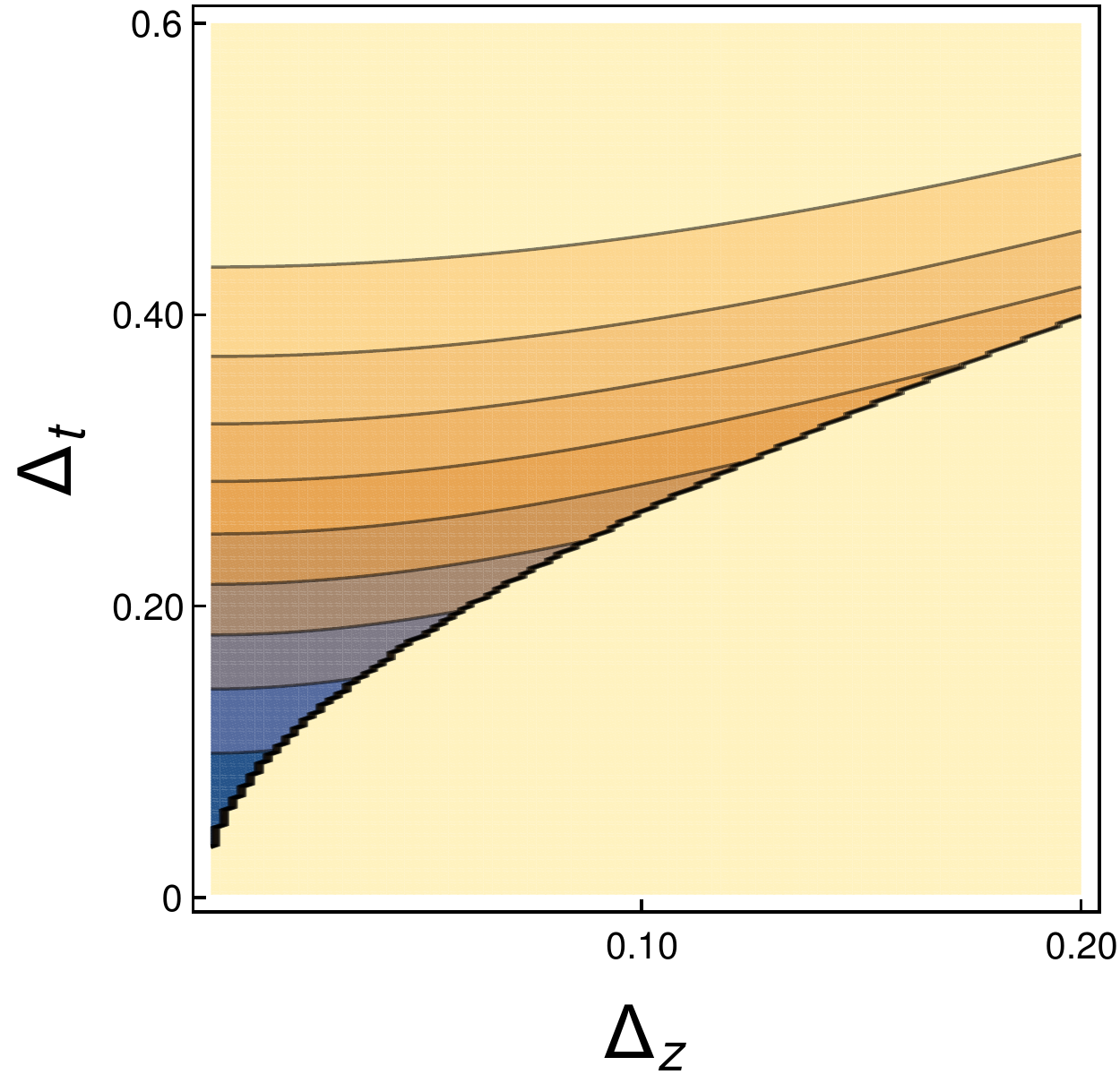} \includegraphics[width=4.2cm]{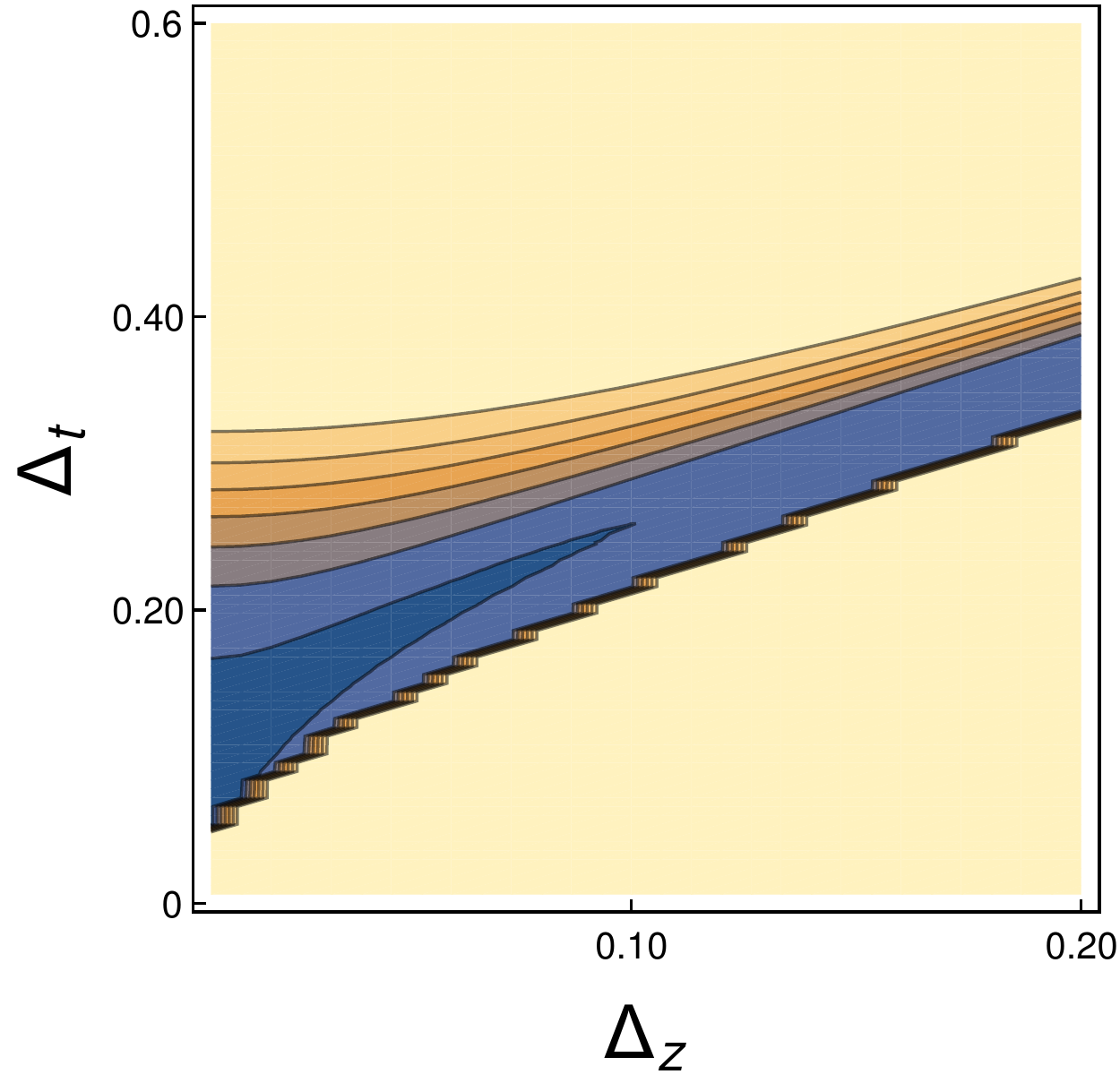}
\end{center}
\caption{Second moment of the Husimi function of the variational ground state as a function of tunneling $\DS$ and Zeeman $\DZ$ gaps  for $\lambda=1$ (left) and $\lambda=3$ (right). Interlayer distance $\delta=\ell_B$. 
We make a 3D plot (top) and a contour-plot (bottom). The canted phase is characterized by the moment valleys (darker zones of the contour-plot), where the wave function is more delocalized. Spin and ppin phases 
are characterized by high moment values (lighter zones of the contour-plot), where the wave function is more localized (coherent). The transition from canted to ppin is better marked for $\lambda=3$ 
that for $\lambda=1$.}\label{MomentDtDZ}
\end{figure}

In figure \ref{MomentDtDZ} we make a 3-dimensional representation and a contour-plot of  $M_{|Z^0_\mathrm{sym}\ra} $ as a function of tunneling $\DS$ and Zeeman $\DZ$ gaps for $\lambda=1$ and $\lambda=3$ 
(we take $\delta=\ell_B$). The figure \ref{MomentDt} corresponds to a cross-section at $\DZ=0.01$. We see as the valley of $M_{|Z^0_\mathrm{sym}\ra}$ (delocalized state), 
represented by darker zones of the contour-plot, captures the canted phase in the $\DS$-$\DZ$ control parameter plane. The transition from canted to ppin phase is better marked (sharp) for $\lambda=3$ than 
for $\lambda=1$.

\subsection{Numeric diagonalization results}

Now we shall diagonalize the Hamiltonian \eqref{Hamlambda} and obtain the corresponding ground state. We shall call it ``exact'' in contrast to the variational ground state discussed in the 
previous section. For zero tunneling $\DS=0$, the Hamiltonian is diagonal in the orthonormal basis \eqref{basisvec}. Its 
eigenvalues can be straightforwardly obtained from \eqref{CCOC} as
\begin{eqnarray}
E_\lambda({}{}^{j,m}_{q_a,q_b})&=&\frac{\varepsilon_\mathrm{cap}(2j+2m-\lambda)^2-8\varepsilon_\mathrm{X}j(j+1)}{2\lambda(2\lambda-1)}\nn\\ 
&&-\frac{\DZ(q_b-q_a)+\Delta_\mathrm{b}(2j+2m-\lambda)}{2\lambda},
\end{eqnarray}
where $\varepsilon_\mathrm{cap}=4\varepsilon_\mathrm{D}-2\varepsilon_\mathrm{X}$ denotes the capacitance energy. 

Looking at $E_\lambda({}{}^{j,m}_{q_a,q_b})$, for small bias $\Delta_\mathrm{b}$, 
the lowest energy state must have zero capacitance energy (note that $\varepsilon_\mathrm{cap}\geq 0$), that is, it must be balanced $2j+2m-\lambda=0$. 
It must also have maximum  angular momentum $j=\lambda/2\Rightarrow m=0$ [remember the constraint $2j+m\leq\lambda$ in \eqref{basisfunc}], which gives the minimum exchange energy. Also, the Zeeman energy attains 
its minimum for $q_b=\lambda/2=-qa$. Therefore, the ground state at $\DS=0$ and small $\Delta_\mathrm{b}$ is the basis state $|\psi_0^s\ra=|{}{}_{-\lambda/2,\lambda/2}^{\lambda/2,\quad 0}\ra$, which coincides with 
the variational CS $|Z^0_\mathrm{sym}\ra=|Z^0_+\ra$ in the spin phase. Actually, the ground state in the spin phase is always $|{}{}_{-\lambda/2,\lambda/2}^{\lambda/2,\quad 0}\ra$, independent of the control parameters 
($\DS,\DZ,\Delta_\mathrm{b}$), and the squared spin expectation value is $\la\psi_0^s|\vec{S}|\psi_0^s\ra^2=\lambda^2$, thus attaining its maximum 
value [remember the identity \eqref{isomag}]. 

For high bias voltage, the dominant part of the energy goes as $-\Delta_\mathrm{b}(2j+2m-\lambda)$ which attains its minimum for $j=0$ and $m=\lambda$ (maximum positive imbalance, i.e. all flux quanta in layer $a$). 
This corresponds to the ppin phase and the ground state in this case is $|\psi_0^p\ra=|{}{}_{0,0}^{0,\lambda}\ra$. This also turns out to be a CS, in fact a particular case of \eqref{u4csfock} given by 
\begin{equation}
|\psi_0^p\ra=|Z_\infty\ra=\frac{\det(\mathbf{a}^\dag)^\lambda|0\ra_\mathrm{F}}{\lambda!\sqrt{\lambda+1}}.
\end{equation}
The squared ppin expectation value is $\la\psi_0^p|\vec{P}|\psi_0^p\ra^2=\lambda^2$, thus attaining its maximum value. 

\begin{figure}[H]
\begin{center}
\includegraphics[width=4.2cm]{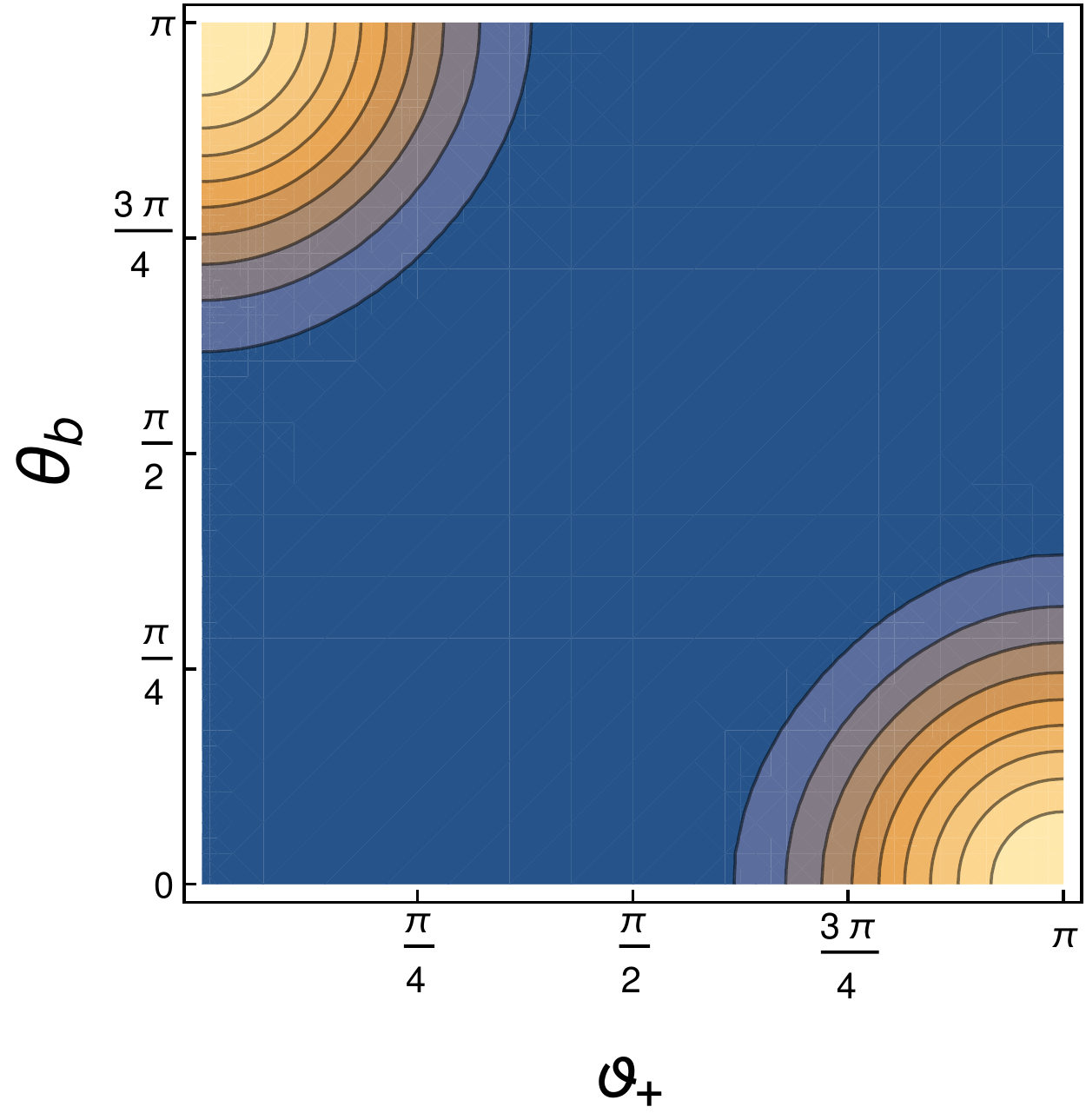} \includegraphics[width=4.2cm]{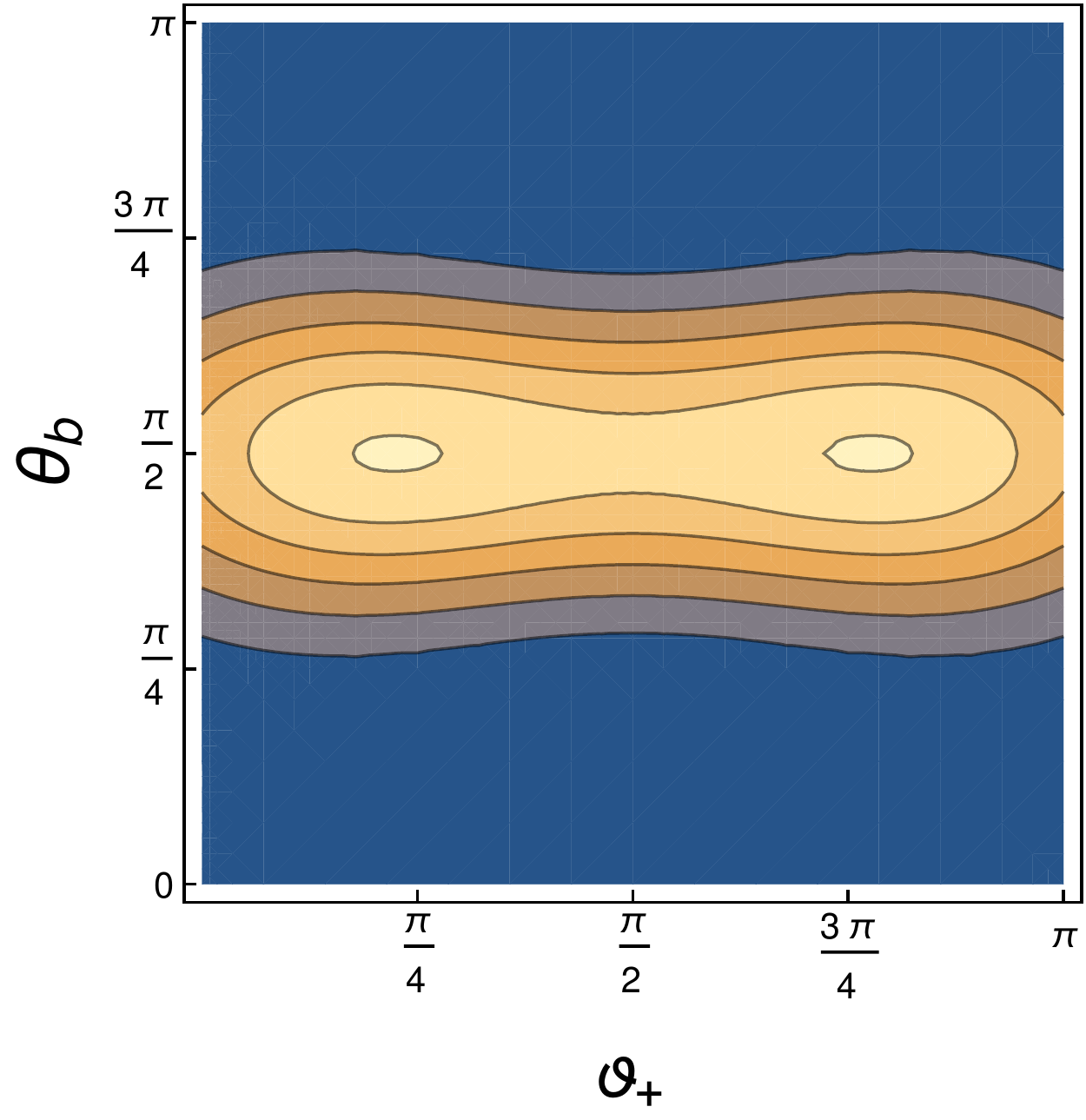}
\includegraphics[width=4.2cm]{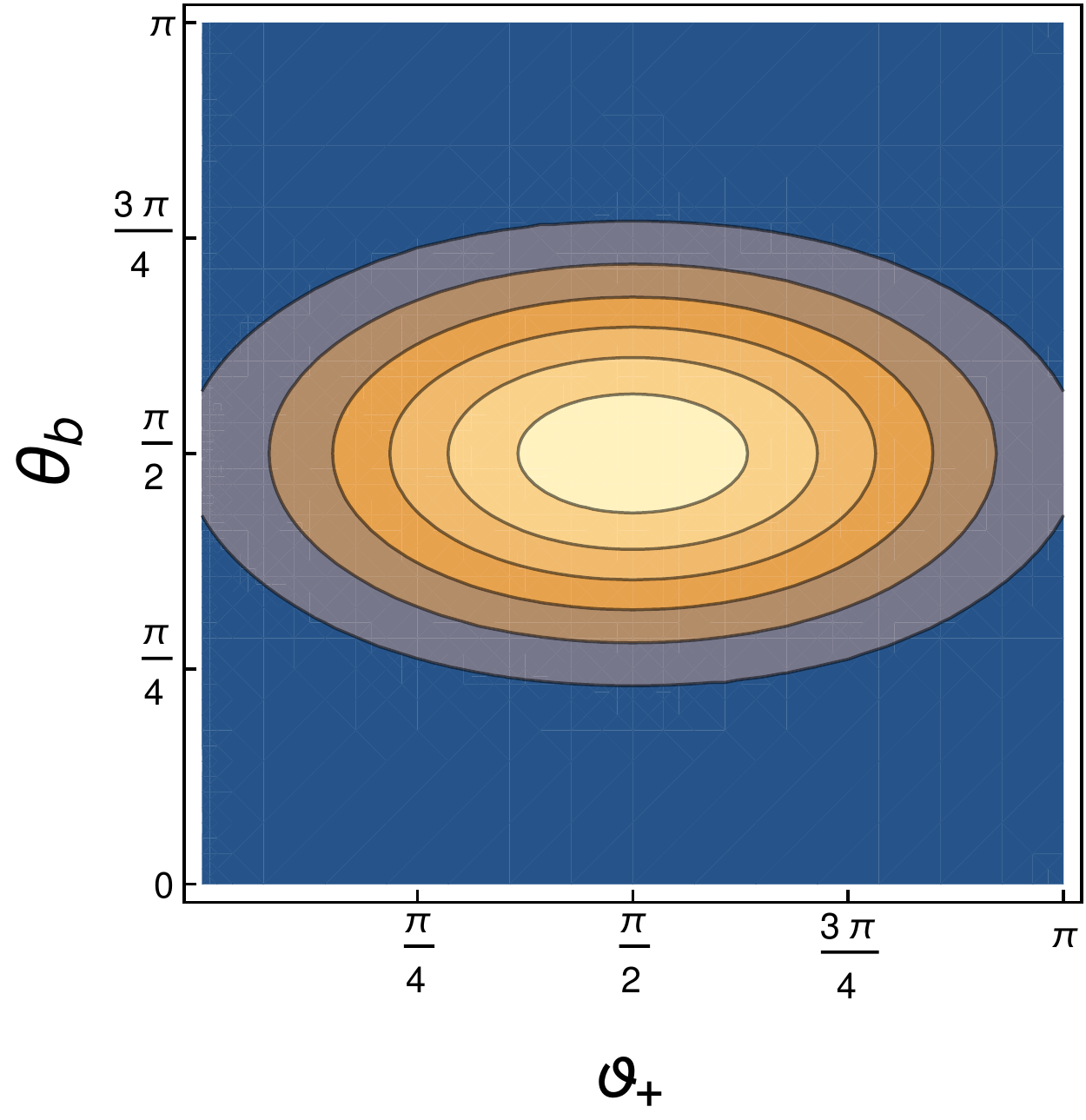} \includegraphics[width=4.2cm]{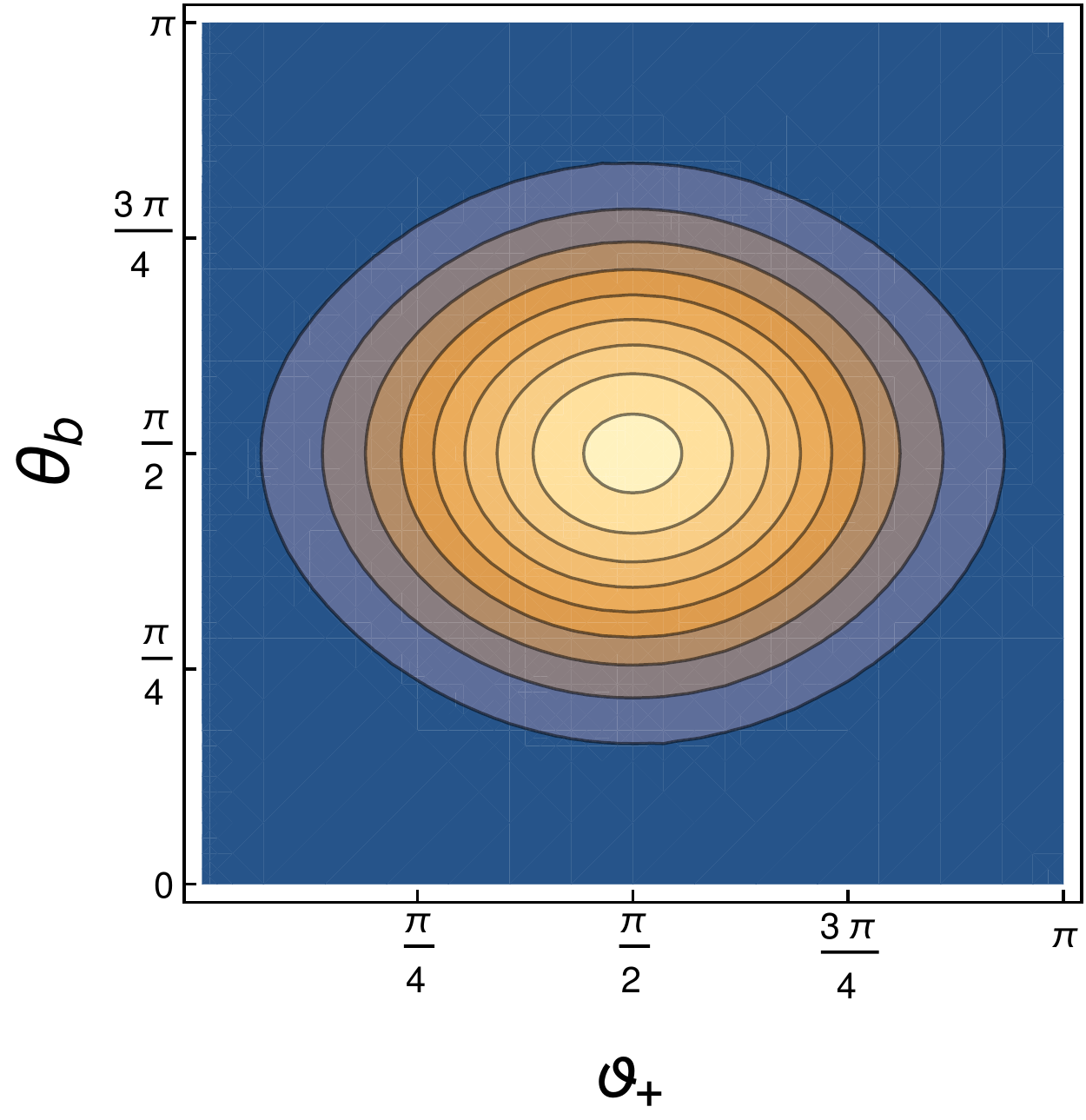}
\end{center}
\caption{Contour plot of the Husimi function $Q_{\psi_0}$ of the exact ground state $|\psi_0\ra$
in the plane  $(\vartheta_+,\theta_b)$ of the phase-space $\mathbb{G}^4_2$ for $\lambda=3$. Same structure and values as in the variational case of figure \ref{contourvar}. }\label{contourq}
\end{figure}

\begin{figure}[h]
\begin{center}
\includegraphics[width=4.2cm]{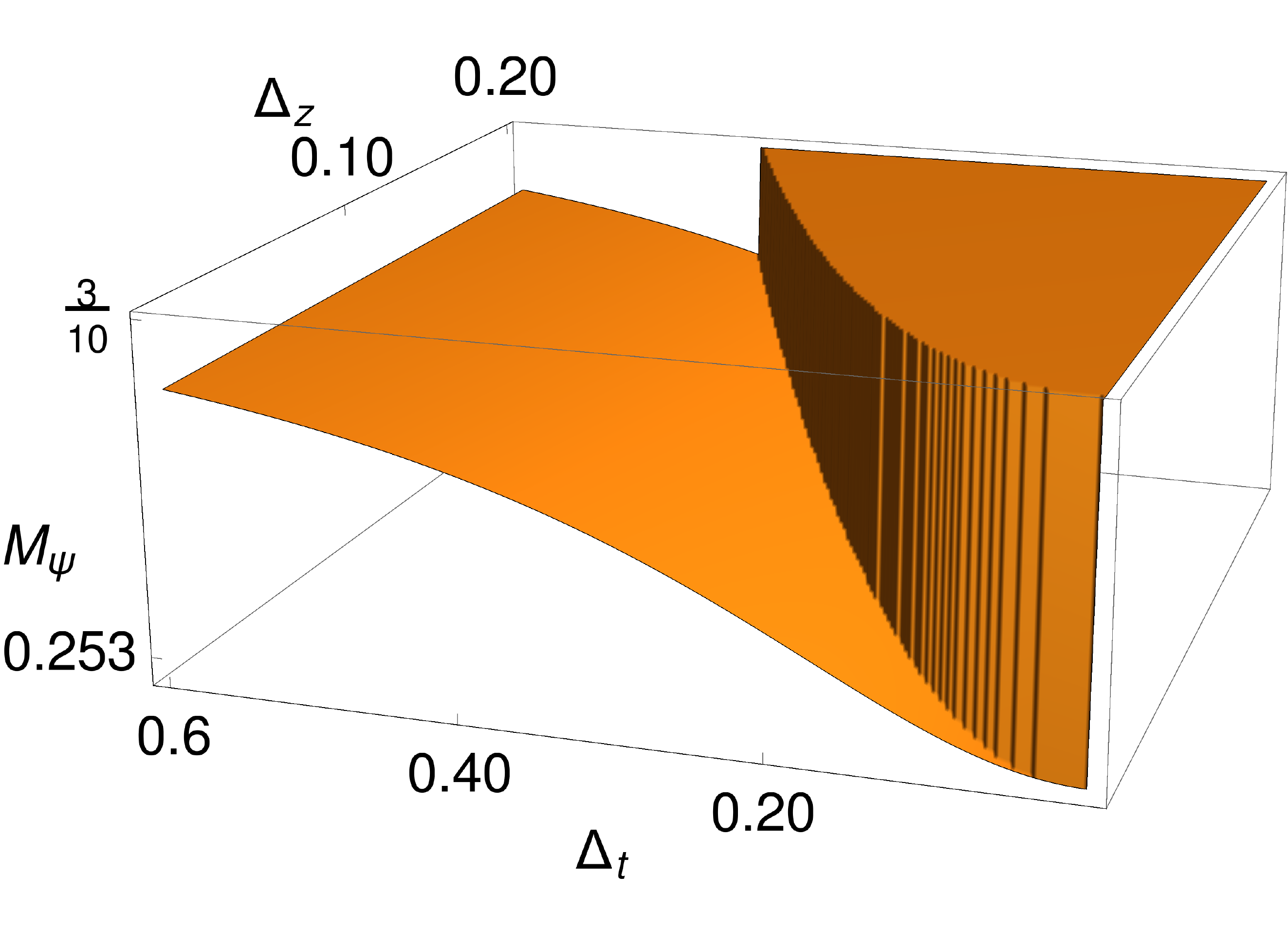} \includegraphics[width=4.2cm]{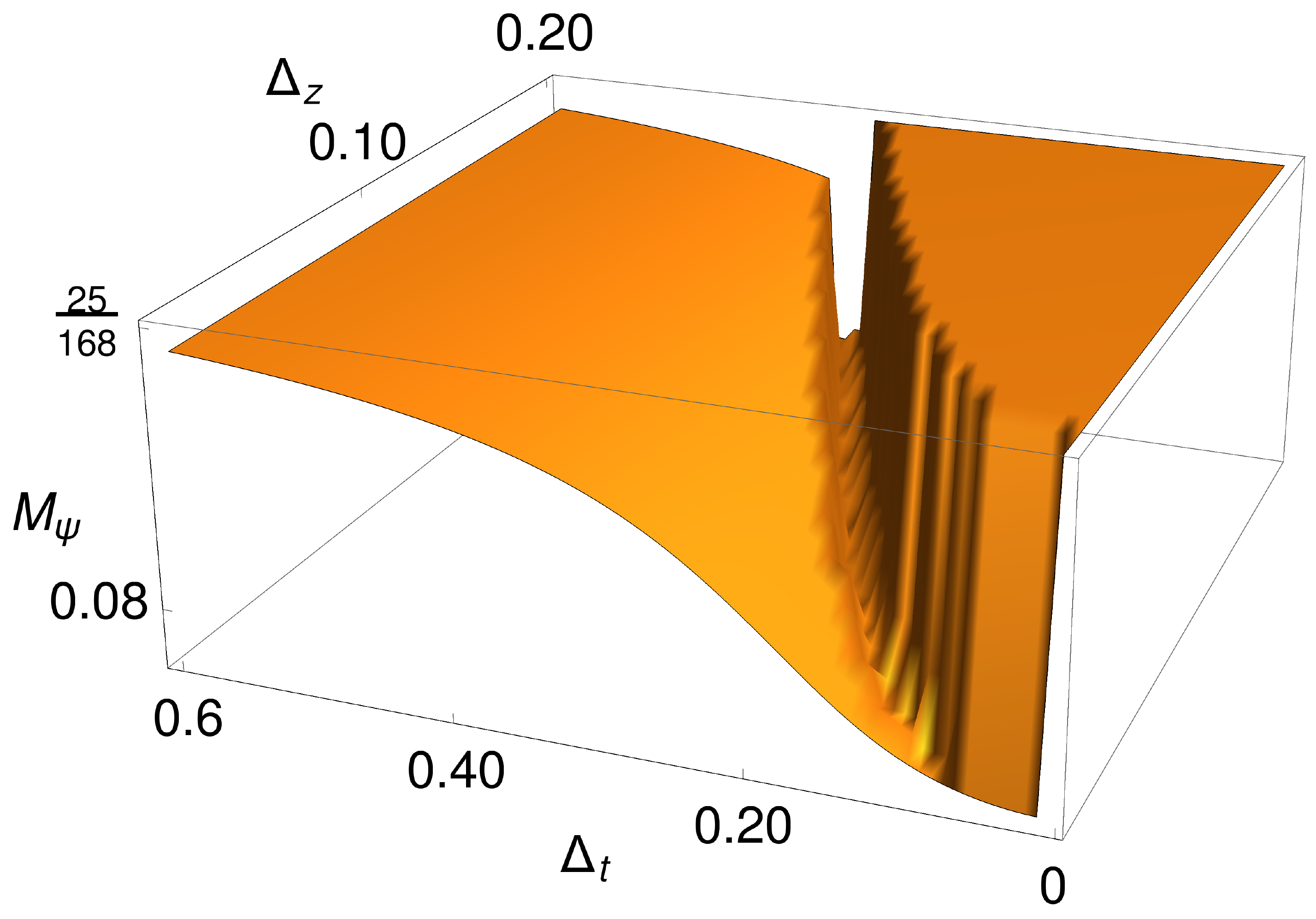}
\\
\includegraphics[width=4.2cm]{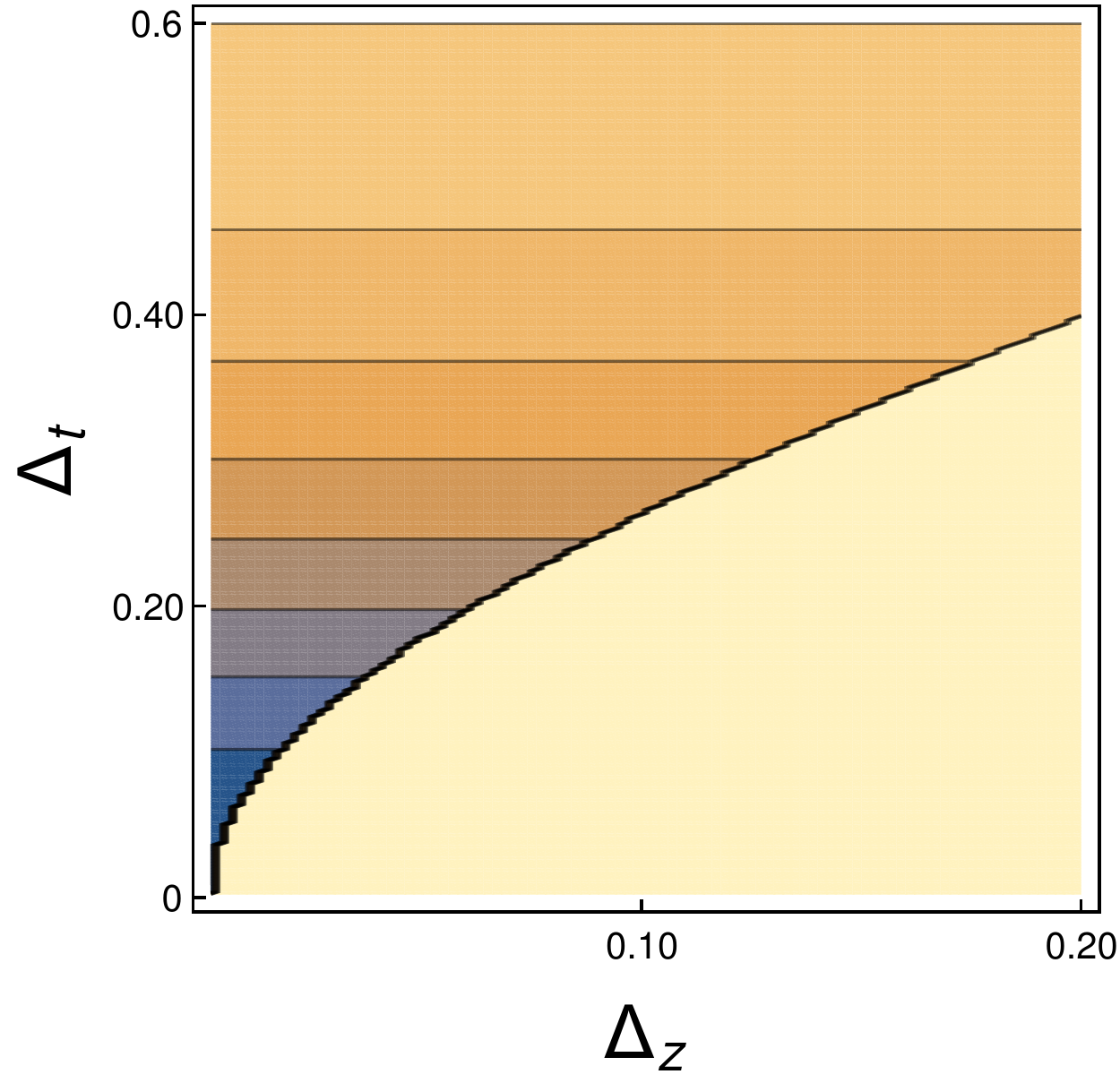} \includegraphics[width=4.2cm]{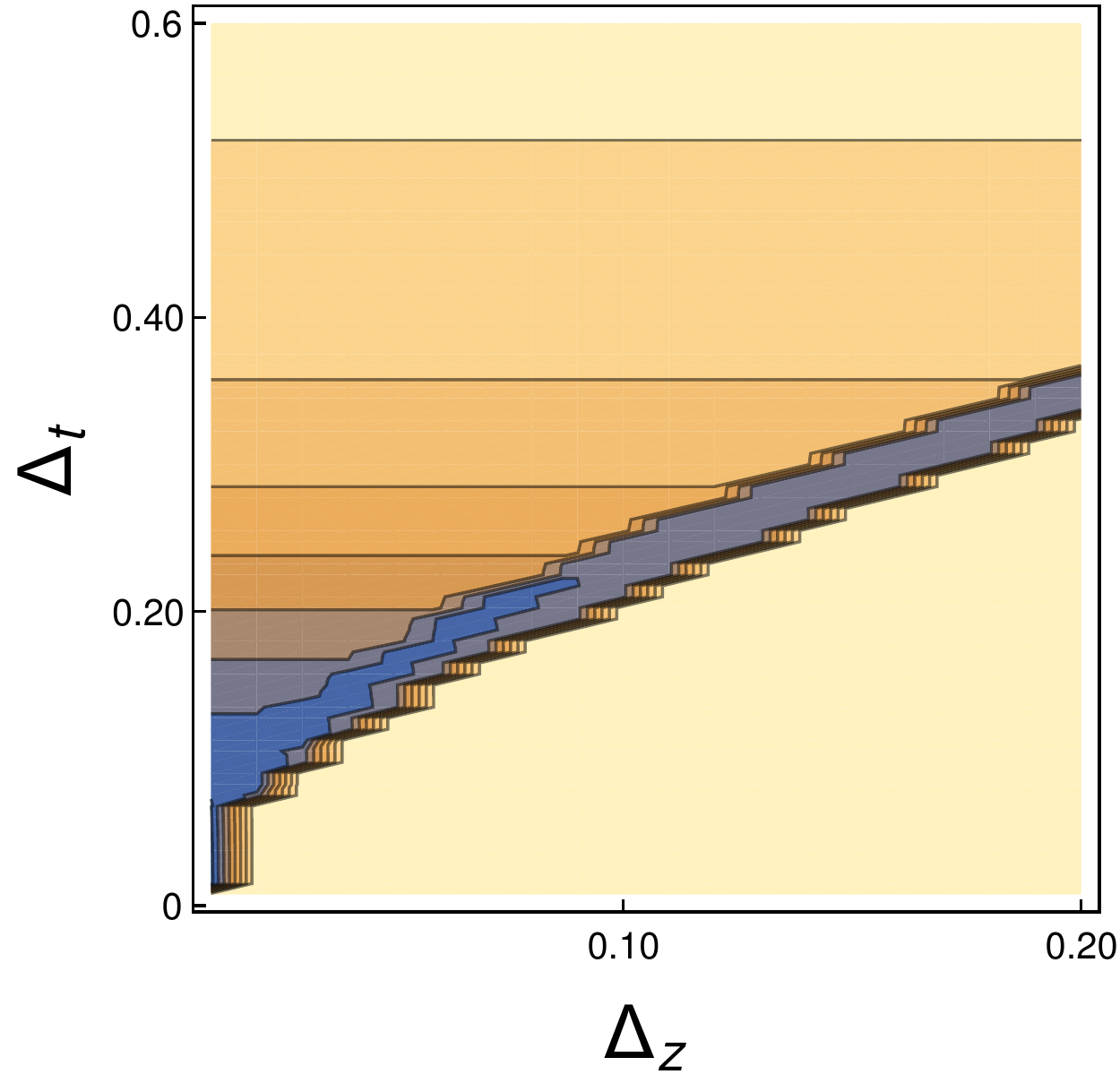}
\end{center}
\caption{Second moment of the Husimi function of the exact ground state as a function of tunneling $\DS$ and Zeeman $\DZ$  for $\lambda=1$ (left) and $\lambda=3$ (right). 
Same structure and values as in the variational case of figure \ref{MomentDtDZ}.}\label{MomentDtDZq}
\end{figure}

For non-zero tunneling, the Hamiltonian \eqref{Hamlambda} is not diagonal in the orthonormal basis \eqref{basisvec}. Indeed, the matrix elements of the interlayer tunneling operator are
\begin{eqnarray}
&&P_1|{}{}_{q_a,q_b}^{j,m}\ra= C_{q_a,q_b}^{j,m+1}|{}{}_{q_a-\um,q_b-\um}^{j-\um,m+1}\ra+
C_{-q_a,-q_b}^{j,m+1}|{}{}_{q_a+\um,q_b+\um}^{j-\um,m+1}\ra+\nn\\ 
&&C_{-q_a+\um,-q_b+\um}^{j+\um,m+2j+2}|{}{}_{q_a-\um,q_b-\um}^{j+\um,m}\ra+
C_{q_a+\um,q_b+\um}^{j+\um,m+2j+2}|{}{}_{q_a+\um,q_b+\um}^{j+\um,m}\ra+\nn\\ 
&&C_{q_a,q_b}^{j,m+2j+1}|{}{}_{q_a-\um,q_b-\um}^{j-\um,m}\ra+
C_{-q_a+\um,-q_b+\um}^{j+\um,m}|{}{}_{q_a-\um,q_b-\um}^{j+\um,m-1}\ra+\nn\\ 
&&C_{-q_a,-q_b}^{j,m+2j+1}|{}{}_{q_a+\um,q_b+\um}^{j-\um,m}\ra+
C_{q_a+\um,q_b+\um}^{j+\um,m}|{}{}_{q_a+\um,q_b+\um}^{j+\um,m-1}\ra,\label{P1coef}
\end{eqnarray}
where the coefficients $C$ where calculated in \cite{GrassCSBLQH} and are given by
\begin{equation}
C_{q_a,q_b}^{j,m}=\frac{1}{2}\frac{\sqrt{(j+q_a)(j+q_b)m(\lambda-(m-2))}}{\sqrt{2j(2j+1)}},\; j\not=0,
\end{equation}
and $C_{q_a,q_b}^{j,m}=0$ for $j=0$. Taking into account the matrix elements \eqref{CCOC} and \eqref{P1coef}, we can calculate the Hamiltonian matrix elements $\la J|H_\lambda|J'\ra$, where 
$J=\{{}{}_{q_a,q_b}^{j,m}\}$ denotes a multi-index running from $J=1,\dots,d_\lambda$. The ground state $|\psi_0\ra$ is a linear combination of the basis states $|J\ra$ as 
\begin{equation}
 |\psi_0(\Delta)\ra=\sum_{J=1}^{d_\lambda} c_J(\Delta)|J\ra,
\end{equation}
with coefficients $c_J(\Delta)$ depending on the Zeeman, tunneling, bias, etc, control parameters (generically denoted by $\Delta$). The Husimi function is then 
\begin{eqnarray}
&Q_{\psi_0(\Delta)}(Z)=|\la Z|\psi_0(\Delta)\ra|^2 &\\ 
 &=\sum_{J,J'=1}^{d_\lambda}  \frac{\overline{\varphi_J(Z)}{\varphi_{J'}(Z)}}{\det(\sigma_0+Z^\dag Z)^{\lambda}} c_J(\Delta)\overline{c_{J'}(\Delta)},&\nonumber
\end{eqnarray}
where $\varphi_J(Z)$ are the homogeneous polynomials \eqref{basisfunc}. The corresponding Husimi function second moment \eqref{MQ} is then given by
\begin{eqnarray}
 &M_{\psi_0}(\Delta)=\sum_{J,J',K,K'=1}^{d_\lambda}c_{J}(\Delta)\overline{c_{J'}(\Delta)}c_{K}(\Delta)\overline{c_{K'}(\Delta)}&\nn\\
 &\times \int_{\mathbb{G}_2^4}\frac{\overline{\varphi_{J}(Z)}{\varphi_{J'}(Z)}\overline{\varphi_{K}(Z)}{\varphi_{K'}(Z)}}{\det(\sigma_0+Z^\dag Z)^{2\lambda}} d\mu(Z,Z^\dag). &
\end{eqnarray}
We have performed a numerical diagonalization of the Hamiltonian  \eqref{Hamlambda} for $\lambda=1$ (dimension $d_1=6$) and $\lambda=3$  (dimension $d_3=50$) using 
a mesh of 300 points, with a resolution of tunneling gap $\DS=0.5$ in figure \ref{MomentDt}, and a mesh of $100\times 30$ points, with a resolution of $\DS=0.6$ and $\DZ=0.2$ in the 3D figure \ref{MomentDtDZq}. 
The multiple integrals in the 8-dimensional Grassmannian ${\mathbb{G}_2^4}$  are also calculated numerically. They are computationally quite hard calculations. 
In figure \ref{MomentDt} we compare variational (dashed curves) with exact (solid curves) values of the Husimi function 
second moment of the ground state. We see that the variational approximation agrees with the exact calculation in the spin phase and captures quite well the delocalization of the ground state in phase space 
inside the canted region. In the ppin region, both the variational and exact grouns states become again localized although, in the exact case, the maximum moment value is only attained asymptotically for high 
$\DS$. This agreement between variational and exact ground states is also patent when comparing figure \ref{contourvar} with \ref{contourq} and figure \ref{MomentDtDZ} with \ref{MomentDtDZq}. The variational result 
captures quite faithfully the ground state structure and localization measures in the three phases.

\begin{figure}[h]
\begin{center}
\includegraphics[width=4.2cm]{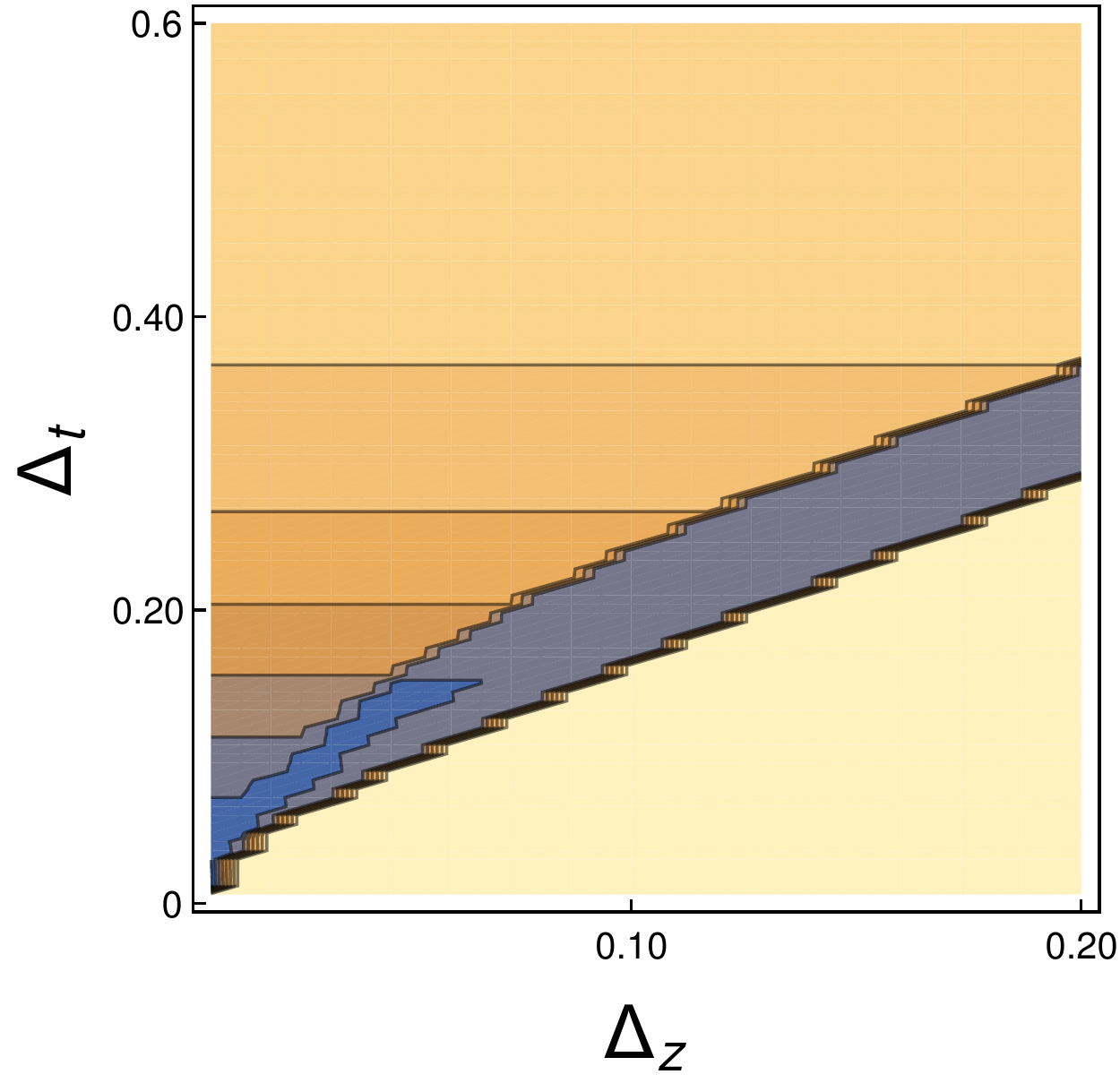} \includegraphics[width=4.2cm]{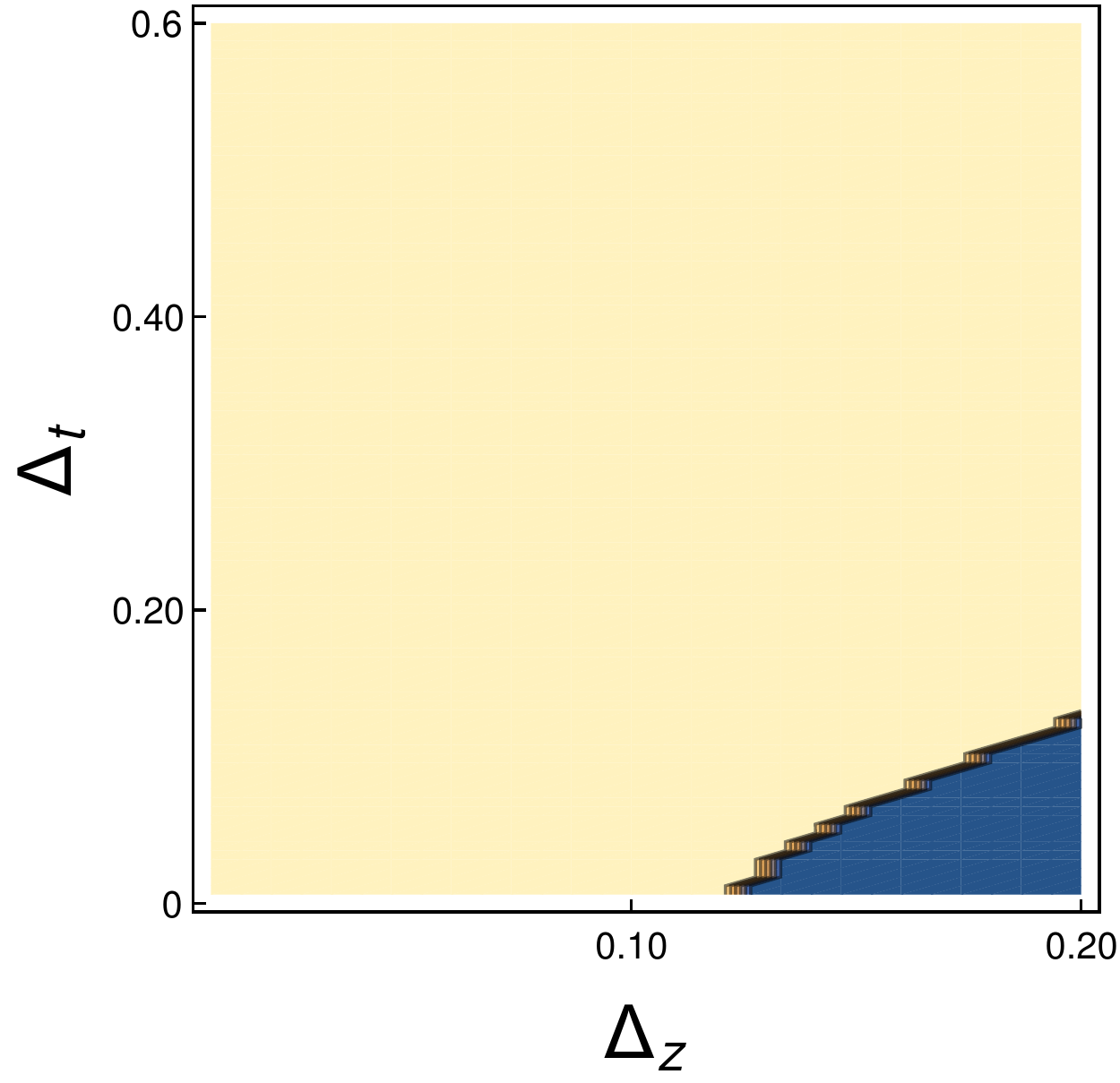}
\end{center}
\caption{Second moment of the Husimi function of the exact ground state as a function of tunneling $\DS$ and Zeeman $\DZ$  for $\lambda=3, \delta=\ell_B$ and bias voltage $\Delta_\mathrm{b}=0.5$ (left) and 
$\Delta_\mathrm{b}=1$ (right). For higher $\Delta_\mathrm{b}$, the ppin phase dominates more and more in the phase diagram $\DZ$-$\DS$. }\label{MomentDtDZqbias}
\end{figure}

In the analytical and variational studies  developed in section \ref{sec3}, we have restricted ourselves to the balanced case, for the sake of simplicity. To finish, and for the sake of completeness, 
we study the effect of a non-zero bias voltage (non-balanced case) on the exact ground state $\psi$  Husimi second moment $M_\psi$.  
In figure \ref{MomentDtDZqbias} we represent contour-plots of $M_\psi$ as a function of tunneling $\DS$ and Zeeman $\DZ$  for $\lambda=3$ and two values of bias voltage: $\Delta_\mathrm{b}=0.5$ and 
$\Delta_\mathrm{b}=1$. We see that a non-zero $\Delta_\mathrm{b}$ modifies the spin-canted and canted-ppin phase transition points as regards the balanced case \eqref{ptp}, here given by transitions from high to low momentum $M_\psi$. 
Therefore, the canted region, characterized by low momentum $M_\psi$ (darker zones in the contour-plot), moves in the phase diagram $\DS$-$\DZ$ when varying $\Delta_\mathrm{b}$. In particular, the second moment 
analysis also reproduces the already noticed fact that the ppin phase dominates at higher values of $\Delta_\mathrm{b}$.

\section{Conclusions and outlook}

Using a coherent state representation of the ground state $\psi$, in the Grassmannian phase space $\mathbb{G}^4_{2}$, given by the Husimi $Q_\psi$ function, 
we have characterized the three quantum phases (spin, ppin and canted) of BLQH system models at 
fractional filling factors $\nu=2/\lambda$. We have found that the Husimi function second moment quantifies the localization (inverse volume) of $\psi$ in phase space and serves as an order parameter distinguishing 
the spin and ppin phases (high localization) from the canted phase (low localization). Otherwise stated, the ground state in spin and ppin phases is highly coherent, whereas in the canted phase it is a kind of 
Schr\"odinger cat, i.e., a superposition of two coherent (quasi-classical, minimal uncertainty) states with negligible overlap. We have also visualized the ground state Husimi function in the spin, ppin and canted phases using two-dimensional cross-sections of the 8-dimensional 
Grassmannian phase space  $\mathbb{G}^4_{2}$. The variational (analytic) treatment produces good qualitative and quantitative results as regards the exact (numeric) diagonalization calculations.

We believe that this CS picture of BLQH systems provides an alternative and useful tool and a new perspective compared to more traditional approaches to the subject.  Indeed, the potentialities  
of the Husimi approach go far beyond the localization analysis of quantum  phases studied in this article. For example, there is possibility of quantum state reconstruction mentioned at the introduction. 
The implementation of these techniques (mainly  imported from quantum optics) in multilayer quantum Hall devices, could open new possibilities for the 
design and use of these nanostructures in quantum informations protocols. Actually, one can find 
quantum computation proposals using BLQH systems in, for example, \cite{Scarola,Yang}. The objective is to engineer quantum Hall states to eventually implement large 
scale quantum computing in multilayer QH systems. For this purpose, controllable (spin and ppin) entanglement \cite{Doucot,Schliemannentang,JPCMenredobicapa}, 
robustness of qubits (long decoherence time and robust interlayer phase difference) \cite{Yang} and easy qubit measurement 
are crucial. For example, in reference \cite{Scarola} it is 
theoretically shown that spontaneously interlayer-coherent BLQH droplets
should allow robust and fault-tolerant pseudospin quantum computation in semiconductor nanostructures. We believe that our BLQH CS $|Z\rangle$ at $\nu=2/\lambda$ will play 
an important role, not only in theoretical considerations but, also in experimental settings.

Moreover, the analysis of small fluctuations 
around the ground state is usually described by a $U(N)$-invariant nonlinear sigma model Lagrangian which, for $N=4$ and filling factor $\nu=2$ acquires the form
\be
L = \tr \left[ (\sigma_0+ Z Z^\dag)^{-1}\partial^\alpha Z  (\sigma_0+ Z^\dag Z)^{-1} \partial_\alpha Z^\dag  \right],
\ee
(plus a Berry phase term) where $Z=z^\mu\sigma_\mu$ is the dynamical $\mathbb G_2^4$ field $Z(x)$ (Goldstone modes) in $(2+1)$ dimensions ($\alpha=0,1,2$). 
This is a generalization of the original Haldane's \cite{HaldanePLA93} description of the continuum field theory describing the low-energy dynamics of the large-spin two-component 
Heisenberg anti-ferromagnet in terms of a $O(3)$-invariant nonlinear sigma model. In fact, this picture 
can be extended to more general $N$-component fractional quantum Hall systems at $\nu=M/\lambda$ and nonlinear sigma models on $\mathbb{G}_M^N$ have already been proposed in  \cite{APsigma}. The structure of the 
Husimi amplitude $\psi(Z)$ of the ground state $\psi$, in each phase, obtained in this article, will be essential to analyze the Goldstone modes describing the small fluctuations around the ground state inside these nonlinear 
sigma models on Grassmannians. This is work in progress.

\section*{Acknowledgements}

The work was supported by the project FIS2014-59386-P (Spanish MINECO and European FEDER funds). C. Pe\'on-Nieto acknowledges the research contract with Ref. 4537 
financed by the  project above.  We all are grateful to Emilio P\'erez-Romero for his valuable collaboration at the early stages of this work.

\end{document}